\documentclass[english,aps,pra,reprint,superscriptaddress,showpacs,showkeys,longbibliography]{revtex4-1}
\usepackage[T1]{fontenc}
\usepackage[utf8]{inputenc}
\usepackage{float,graphicx}
\makeatletter
\usepackage{amsmath,amssymb,amstext,color,braket,bm}
\usepackage{txfonts,comment,dcolumn,wasysym}
\usepackage{booktabs,changes}

\definecolor{mygreen}{rgb}{0,0.5,0}
\definecolor{myblue}{rgb}{0,0,0.75}
\definecolor{mymagenta}{cmyk}{0,1,0,0.12}

\usepackage[colorlinks,bookmarks=false,citecolor=blue,linkcolor=red,urlcolor=blue]{hyperref}

\def\wt#1{\inner(#1)}
\def\inner(#1,#2,#3,#4,#5,#6){\ensuremath\left(\begin{array}{ccc} #1 	& #2 & #3 \\ #4 & #5 & #6 \end{array}\right)}
\def\ws#1{\innerv(#1)}
\def\innerv(#1,#2,#3,#4,#5,#6){\ensuremath\left\{\begin{array}{ccc} #1 & #2 & #3 \\ #4 & #5 & #6 \end{array}\right\}}

\newcommand\minus{
  \setbox0=\hbox{-}
  \vcenter{
    \hrule width\wd0 height \the\fontdimen8\textfont3
  }%
}

\def\pone{\tfrac{1}{2}}
\def\mone{\minus\tfrac{1}{2}}
\def\pthree{\tfrac{3}{2}}
\def\mthree{\minus\tfrac{3}{2}}
\newcommand{\sgn}{\operatorname{sgn}}
\makeatother
\usepackage{babel}

\begin{document}
\title{{\em Magic distances} in the blockade mechanism of Rydberg $p$ and $d$ states}

\author{B.~Vermersch}
\email{benoit.vermersch@uibk.ac.at}
\affiliation{Institute for Quantum Optics and Quantum Information of the Austrian
Academy of Sciences, A-6020 Innsbruck, Austria}
\affiliation{Institute for Theoretical Physics, University of Innsbruck, A-6020
Innsbruck, Austria}

\author{A.~W.~Glaetzle}
\affiliation{Institute for Quantum Optics and Quantum Information of the Austrian
Academy of Sciences, A-6020 Innsbruck, Austria}
\affiliation{Institute for Theoretical Physics, University of Innsbruck, A-6020
Innsbruck, Austria}

\author{P.~Zoller}
\affiliation{Institute for Quantum Optics and Quantum Information of the Austrian
Academy of Sciences, A-6020 Innsbruck, Austria}
\affiliation{Institute for Theoretical Physics, University of Innsbruck, A-6020
Innsbruck, Austria}

\date{\today}
\begin{abstract}
We show that the Rydberg blockade mechanism, which is well known in the case of $s$ states, can be significantly different for $p$ and $d$ states due to the van der Waals couplings between different Rydberg Zeeman sublevels and the presence of a magnetic-field. We show, in particular, the existence of magic distances  corresponding to the laser-excitation of a superposition of doubly excited states.
\end{abstract}
\maketitle

\section{Introduction\label{sec:intro}}
Rydberg states of alkali atoms have remarkable properties~\cite{gallagher2005rydberg} including long lifetimes and very high polarizabilities which make them ideally suited to study a plethora of quantum phenomena. Due to the long-range nature of their interactions Rydberg atoms allow, in particular, to study various spin models \cite{Low2012} and also have applications for quantum information processing~\cite{Saffman2010}.

A central property of a laser-excited Rydberg gas is the \emph{blockade} mechanism \cite{Jaksch2000}: Two non-interacting ground-state atoms can be simultaneously excited with a coherent laser to a Rydberg level. However, when the distance between the atoms becomes smaller than a typical length scale, the blockade radius, the doubly excited state is shifted out of resonance due to the strong interactions between Rydberg levels. This mechanism can be generalized to the case of arbitrarily large number of atoms \cite{Lukin2001} leading to an effective two-level system -- \emph{the super-atom} -- consisting of the state with all atoms in their ground state and the symmetric state with a single Rydberg excitation shared between all atoms. They are coupled with an enhanced Rabi frequency which has been successfully demonstrated in recent experiments \cite{Tong2004,Singer2004,Gaetan2009,Urban2009,Ebert2014}.

In typical experimental setups with optical lattices or dipole traps \cite{Ebert2014,Barredo2014,Schauss2014,Hankin2014,Viteau2011}, Rydberg atoms interact via van der Waals (vdW) interactions scaling as $\propto 1/r^6$, with $r$ the interparticle distance. For the familiar case of Rydberg $s$ states these interactions are isotropic which result in a spherical symmetric blockade region. In contrast, for higher angular momentum states, e.g., Rydberg $p$ or $d$ states, these interactions can be strongly anisotropic~\cite{Reinhard2007} considering a specific Zeeman sublevel only. 

However, in general, the vdW interactions also mix populations in different Zeeman sublevels~\cite{Walker2008} and the blockade mechanism cannot simply be described by treating the atoms as two-level systems consisting of a ground and a single Rydberg state. For Rydberg $s$ states the strength of the vdW mixing matrix element is much smaller than the diagonal interaction matrix element and this treatment, considering only a single Zeeman sublevel,  is sufficient. On the contrary, for Rydberg $p$ and $d$ states the vdW mixing matrix element is of the same order as the diagonal interaction matrix element leading to much more complex blockade dynamics which we will discuss in the following.

Complementary to recent work on spin-spin interactions between ground-state atoms with laser-admixed Rydberg interactions involving $p$ states~\cite{Glaetzle2014}, we are interested here in the dynamics of an ensemble of atoms, which are excited, \emph{resonantly} with the bare atomic transition, to a $p$ (or $d$) Rydberg state, as in a typical Rydberg blockade experiment. The goal of the paper is to give a general theoretical description of the blockade mechanism taking into account the vdW couplings between Zeeman levels but also the {\em anisotropic} character of the vdW interactions and the influence of a magnetic-field. We show, in particular, that the interplay between the magnetic-field and the vdW interactions leads to the existence of \emph{magic distances} where the laser can resonantly excite a superposition of doubly excited Rydberg states -- even inside the blockade radius. Furthermore, our general treatment allows us to interpret a recent experiment demonstrating the anisotropic Rydberg blockade using $D_{3/2}$ states~\cite{Barredo2014}.

Our paper is organized as follows: In Sec.~\ref{sec:method} we illustrate the Rydberg blockade in the presence of vdW coupling for the simplest possible example of two Rydberg $P_{1/2}$ states. In Sec.~\ref{sec:dstates} we generalize our approach to Rydberg $P_{3/2}$ and $D_{3/2}$ states and provide a theoretical interpretation of the recent experimental results of Barredo {\it et al.}~\cite{Barredo2014}. Furthermore, we propose parameters for a {\em magic distance} experiment with Rydberg $D_{3/2}$ states. Finally, in Sec.~\ref{sub:magic} we describe the consequences of the presence of a magic distance at the many-body level and generalize the concept of a super-atom. In Appendix~\ref{app:vdW} we review the properties of the anisotropic vdW interactions whereas in Appendix~\ref{app:quad} we assess the effect of quadrupole-quadrupole interactions.

\section{Generalized Rydberg blockade in the $P_{1/2}$ manifold\label{sec:method}}
\begin{figure}[tb]
\begin{centering}
\includegraphics[width=0.99\columnwidth]{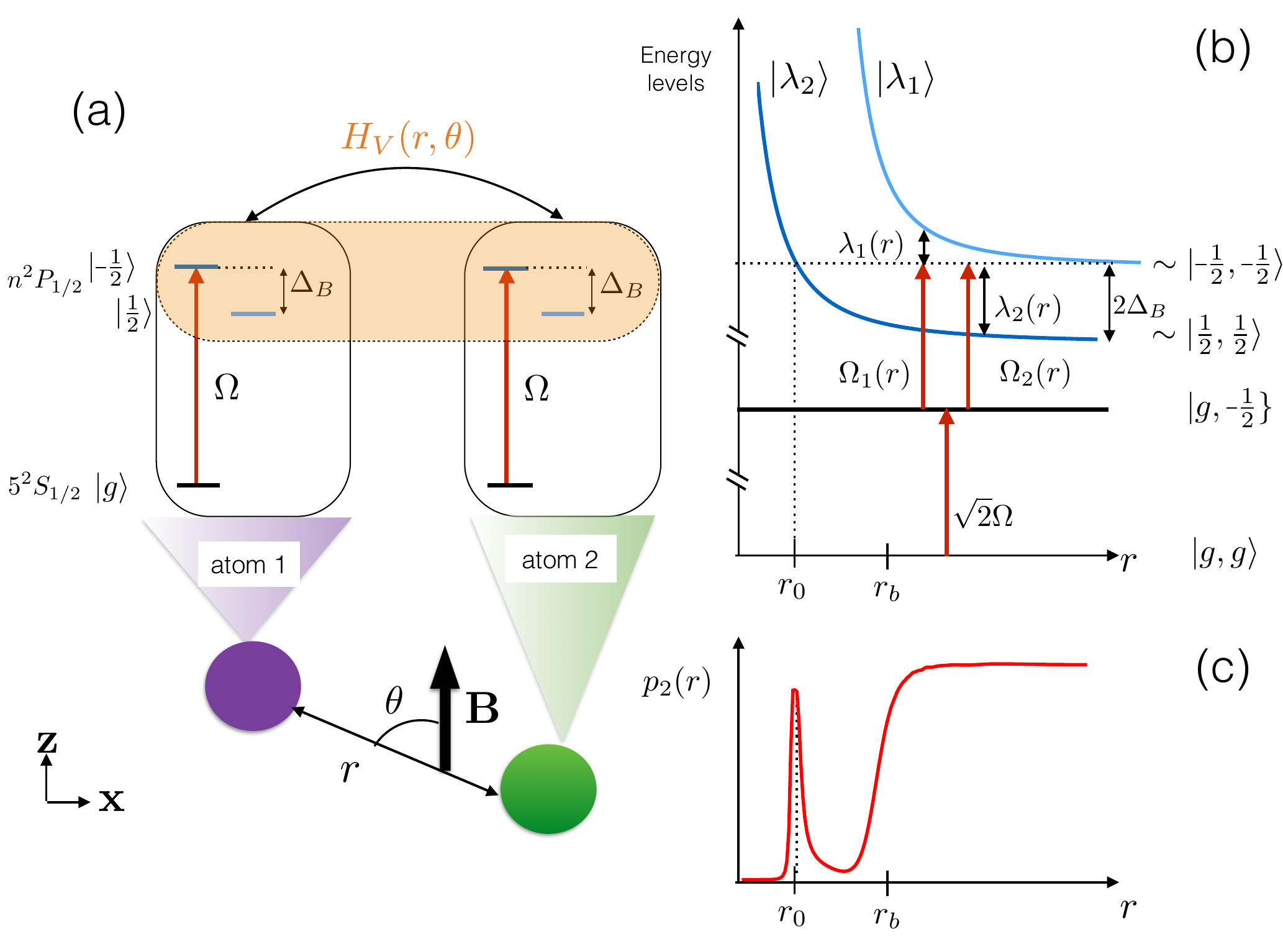}
\par\end{centering}
\caption{
Setup: (a) Two atoms (green and violet circles), separated by a distance $r$ and an angle $\theta$ with respect to the magnetic-field {\bf B} (thick black arrow), are laser-excited to the \mbox{$|n^2P_{1/2},m_j=\mone\rangle$} Rydberg states which interact via vdW interactions $H_V(r,\theta)$ of Eq.~\eqref{eq:H_V}.
(b) Principle of the generalized Rydberg blockade for the simplest case $\theta=\pi/2$: vdW interactions will mix the Rydberg states  $|\pone,\pone\rangle$ and $|\mone,\mone\rangle$ yielding new eigenstates $|\lambda_1\rangle$ and $|\lambda_2\rangle$ which can be both laser-excited from the symmetric state \mbox{$|g,\mone\}=\left(|g,\mone\rangle+|\mone,g\rangle\right)/\sqrt{2}$} with Rabi frequencies $\Omega_1(r)$ and $\Omega_2(r)$ (red arrows), respectively, where the position dependency reflects the spatially dependent populations of  $|\pone,\pone\rangle$ and $|\mone,\mone\rangle$ in $|\lambda_i\rangle$. In particular, for repulsive vdW interactions and Zeeman splitting $\Delta_B>0$ both atoms can be resonantly excited to $|\lambda_2\rangle$ at the magic distance $r_0$, whereas in the limits $r\to 0$ and $r\to\infty$, we obtain the standard blocked and non-interacting situation, respectively. (c) Sketch of the average  probability of double excitation $p_2(r)$ in presence of a magic distance.
}
\label{fig:setup}
\end{figure}

In this section we illustrate the generalized blockade picture discussing the conceptually simplest example 
of two atoms laser-excited to a specific Zeeman sublevel in the Rydberg $P_{1/2}$ manifold [see Fig.~\ref{fig:setup} (a)]. In this case the vdW interaction Hamiltonian, $H_V$, takes a simple form, which allows us to analytically describe the dynamics of the system beyond the familiar Rydberg $S_{1/2}$ state picture \cite{Saffman2010}. Figure~\ref{fig:setup}(b) schematically illustrates the new features of the generalized Rydberg blockade mechanism:  At large distance, $r\gg r_b$, where vdW interactions are negligibly small, two atoms in the atomic ground-state $|g,g\rangle$ can be resonantly laser-excited to a Rydberg state $|\mone,\mone\rangle$. At smaller distances, $r\sim r_b$,  the doubly excited Rydberg state $|\mone,\mone\rangle$ will in general shift out of resonance because of the strong vdW interaction resulting in the well-known Rydberg blockade effect~\cite{Saffman2010}.  However, due to the vdW couplings between doubly excited Rydberg states, e.g.  for $\theta=\pi/2$ between $|\mone,\mone\rangle$ and another Rydberg state $|\pone,\pone\rangle$, new eigenstates $|\lambda_1\rangle$ and $|\lambda_2\rangle$ are formed which asymptotically (for large distances) connect to the bare states $|\pone,\pone\rangle$ and $|\mone,\mone\rangle$, respectively. In the presence of an external magnetic-field the latter two states are Zeeman split in energy by $2\hbar\Delta_B$. In the generalized blockade picture (including vdW couplings) both states $|\lambda_1\rangle$ and $|\lambda_2\rangle$ contain a contribution of $|\mone,\mone\rangle$ and thus can be laser-excited from the ground-state with detunings $\lambda_1(r)$ and $\lambda_2(r)$, respectively \footnote{These quantities can be interpreted as the Born-Oppenheimer energies of the vdW interactions}. In the particular configuration of Fig.~\ref{fig:setup}(b) where $\Delta_B>0$ the state $|\lambda_2\rangle$ shifts into resonance at a {\em magic distance} $r_0$ and can be resonantly laser-excited even within the blockade radius $r_b$. This will lead to a peak in the average probability of double excitation $p_2(r)$ sketched in panel~(c). In the following section we will derive the position, height and width of this additional peak analytically for $nP_{1/2}$ Rydberg states. In Sec.~\ref{sec:dstates} we generalize our approach to Rydberg $P_{3/2}$ and $D_{3/2}$ where several Zeeman sublevels are coupled resulting in a multi-peak structure of $p_2(r)$ shown, for example, in Fig.~\ref{fig:Vvsrtheta}.

\subsection{Description of the system}
We consider the setup shown in Fig.~\ref{fig:setup}. Two alkali atoms located at positions $\bm{r}_i$ ($i=1,2$) are laser-excited to a particular Rydberg state in the $nP_{1/2}$ manifold, which interact via vdW interactions $H_V(\bm{r})$, where $\bm{r}=\bm{r}_2-\bm{r}_1$.
Our goal is to study the competition between the laser excitation and the vdW interactions as in a typical Rydberg blockade situation. As a particular example we consider Rubidium atoms initially in their electronic ground state, which we choose as \mbox{$|g\rangle\equiv|5^2S_{1/2},F=2,m_F=2\rangle_z=|5^2S_{1/2},m_j=\pone\rangle_z\otimes|N\rangle$}. Here, \mbox{$|N\rangle=|I=\pthree,m_I=\pthree\rangle_z$} denotes the nuclear spin state. A resonant laser with Rabi frequency $\Omega$ excites the atoms to one of the \mbox{$|m_j=\pm\pone\rangle\equiv|n^2P_{1/2},m_j=\pm\pone\rangle_z\otimes|N\rangle$} Rydberg states \footnote{Given their negligible hyperfine splitting \cite{Saffman2010,Low2012}, we can describe the Rydberg states using the basis associated to the fine-structure, thus considering that the nuclear spin behaves as a spectator.}. The presence of the magnetic-field $\bm{B}=B\bm{z}$ gives rise to a Zeeman splitting of the Rydberg energy levels described by the Hamiltonian
\begin{equation}
H_B^{(i)} = \sum_{m_j}\left(\hbar\omega_{nP_{1/2}}+ \Delta E_{m_j}-  \Delta E_g \right)|m_j\rangle_i\langle m_j|,
\end{equation}
where $\hbar \omega_{nP_{1/2}}$ is the energy of the transition between the ground state $|g\rangle$ and the Rydberg states $|m_j\rangle$ in the absence of magnetic field. With $\Delta E_{m_j}\equiv\mu_Bg_j Bm_j $, we denote the Rydberg Zeeman shift where $\mu_B = 1.4\,h\,$MHz/G is the Bohr magneton and $g_j$ is the Land\'e Factor for the Rydberg level. Finally, $\Delta E_{g}\equiv\mu_Bg_F Bm_F $, with $g_F$ the ground state Land\'e Factor, is the ground-state Zeeman shift. We note that the value of the magnetic field is chosen sufficiently small to neglect higher order Zeeman effects. Furthermore, the Zeeman shifts $\Delta E_{m_j}$ should be much smaller than the fine structure splitting $ \Delta_{\mathrm{fs}}$ (typically several GHz for $n=30-40$) to neglect couplings between fine structure manifolds (Paschen-Back effect).

We now consider excitation of the Rydberg level $|\mone\rangle$ from the ground-state $|g\rangle$ with a laser propagating along the direction of the magnetic-field $\bm{k_L}=k_L\bm{z}$ with left circular polarization. In the rotating wave approximation, the resulting Hamiltonian is 
\begin{equation}
H_L^{(i)} = \frac{\hbar \Omega}{2} \left[e^{i(\bm{k_L}.\bm{r_i}-\omega_L t)} |g\rangle_i\langle \mone|  +\mathrm{hc}\right],
\end{equation}
with $\omega_L$ the laser frequency and \mbox{$\Omega=-\langle g|\bm{\varepsilon}.\bm{d}|\mone\rangle \mathcal{E}/\hbar$} the Rabi frequency. Here $\mathcal{E}$  is the electric-field field amplitude, $\bm\epsilon$ its polarization ($\sigma^-$ here) and $\bm{d}$ is the dipole operator. In the rotating frame and absorbing the phase $e^{i\bm{k_L}.\bm{r_i}}$ term in the definition of $|g\rangle_i$, the single-particle Hamiltonian $H_A^{(i)}=H_L^{(i)}+H_B^{(i)}$ reduces to 
\begin{equation}
H_A^{(i)} = \frac {\hbar \Omega}{2} \left[|g\rangle_i\langle \mone|+ |\mone\rangle_i\langle g|\right] -\hbar\Delta_B|\pone\rangle_i\langle\pone|,
\end{equation}
where the laser frequency $\omega_L=\omega_{nP_{1/2}} + (\Delta E_{-1/2} - \Delta E_g)/\hbar$ is chosen to be in resonance with the atomic transition and the Zeeman splitting is  \mbox{$\hbar\Delta_B=\Delta E_{-1/2}-\Delta E_{1/2}$}.

\begin{figure}[t]
\centering{}
\includegraphics[width=0.45\textwidth]{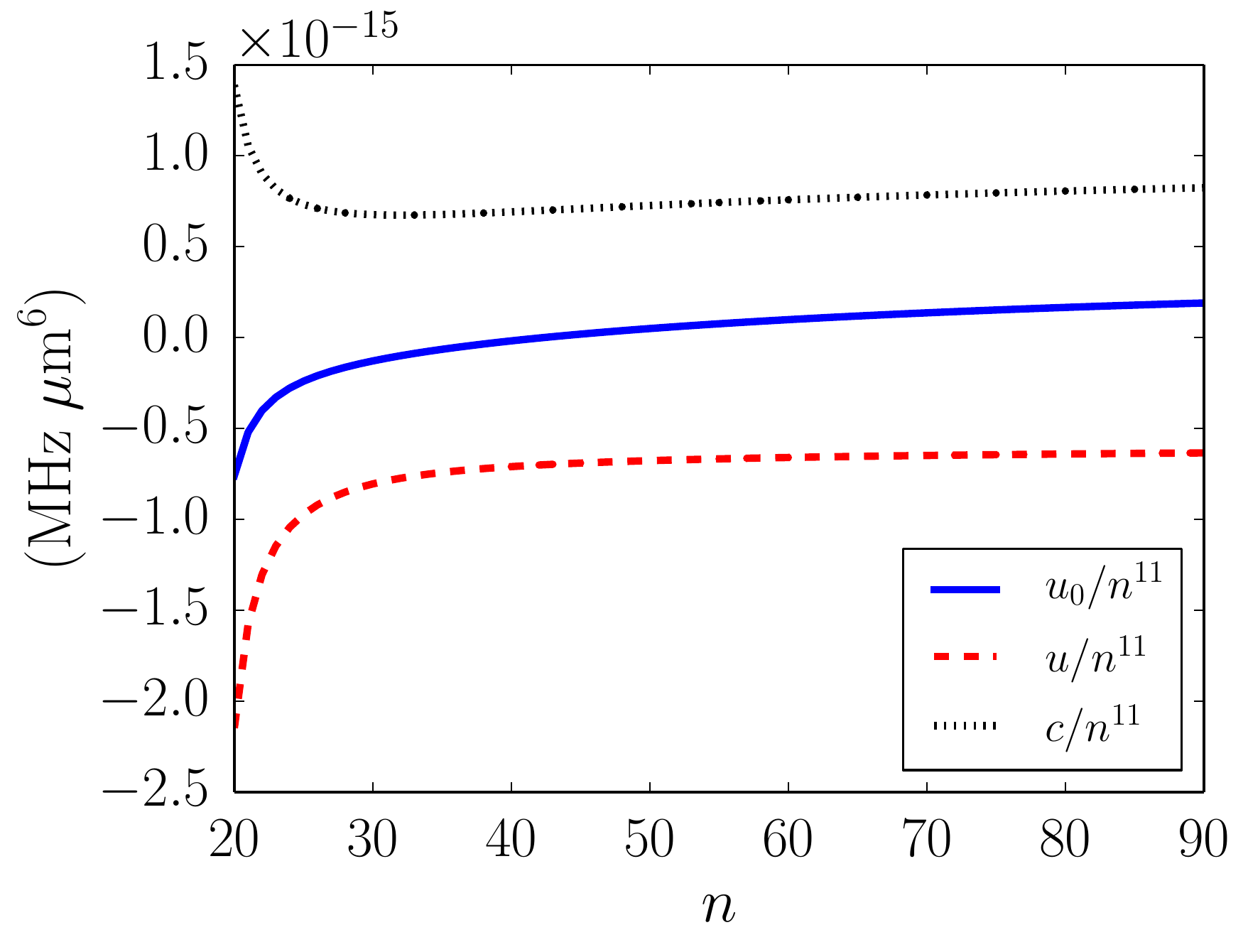}
\caption{$u_0$, $u$ and $c$ as a function of the principal quantum number $n$ for Rubidium $P_{1/2}$ states. In contrast to $S_{1/2}$ states, the coupling term has the same order of magnitude as the diagonal terms for $P_{1/2}$ states. \label{fig:uc}}
\end{figure}

The two Rydberg atoms interact dominantly via dipole-dipole interactions. At large distances and away from F\"orster resonances \cite{Saffman2010} these interactions can be treated perturbatively leading to the vdW Hamiltonian $H_V$ \cite{Walker2008,Glaetzle2014}. We describe in detail the derivation of $H_V$ in the various Rydberg fine-structure manifolds in Appendix~\ref{app:vdW}. Here, we are interested in $nP_{1/2}$ Rydberg states where 
\begin{equation}
H_V(\bm{r}) = [u_0 \mathbb{I}+c\mathcal{D}(\theta)]/r^6
\end{equation}
consists of a diagonal, isotropic, part, with  $\mathbb{I}$ the $4\times 4$ identity matrix, and an anisotropic, non-diagonal, part
\begin{equation}
\mathcal{D}(\theta)=\left[\begin{matrix}- \sin^{2} {\theta} & - \frac{1}{2} \sin {2 \theta} & - \frac{1}{2} \sin {2 \theta} & \sin^{2} {\theta}\\- \frac{1}{2} \sin {2 \theta} & \sin^{2} {\theta} - \frac{2}{3} & \sin^{2} {\theta} - \frac{4}{3} & \frac{1}{2} \sin {2 \theta}\\- \frac{1}{2} \sin {2 \theta} & \sin^{2} {\theta} - \frac{4}{3} & \sin^{2} {\theta} - \frac{2}{3} & \frac{1}{2} \sin {2 \theta}\\\sin^{2} {\theta} & \frac{1}{2} \sin {2 \theta} & \frac{1}{2} \sin {2 \theta} & - \sin^{2} {\theta}\end{matrix}\right],\label{eq:H_V}
\end{equation}
written here in the basis $(|\mone,\mone\rangle,|\mone,\pone\rangle,|\pone,\mone\rangle,|\pone,\pone\rangle)$ with the notation \mbox{$|m_1,m_2\rangle\equiv |m_1\rangle_1\otimes|m_2\rangle_2$}. Here, $r=|\bm{r}_1-\bm{r}_2|$ and \mbox{$\theta=\angle(\bm{r},\bm{B})$} is the angle between the relative vector and the magnetic field. The prefactors $u_0$, $c$ [given by Eq.~\eqref{eq:u0} and \eqref{eq:c}, respectively] are generalized vdW coefficients. We note that this interaction Hamiltonian neglects the couplings to the neighboring $P_{3/2}$ Rydberg manifold and is therefore only valid for distances $r$ larger than a typical length scale $r^\star=(\Delta_{\mathrm{fs}}/u_0)^{1/6}$. The first term of~\eqref{eq:H_V} corresponds to the diagonal part of the interactions, and does not depend on the angle $\theta$, thus representing the typical and well-known vdW potential $\propto 1/r^6$. The second term includes anisotropic diagonal matrix elements as well as couplings between the different $m_j$ levels. The values of $u_0$ and $c$ are shown in Fig.~\ref{fig:uc} for Rubidium $P_{1/2}$ states. They are of the same order of magnitude so that one can expect the anisotropic characters of the interactions and the vdW couplings to have a significant impact on the dynamics. We note that considering $S_{1/2}$ states, $H_V$ can be written in the same form as \eqref{eq:H_V}. However, the existence of the coupling term $c$ is in this case only due to the fine-structure splitting of the neighboring $p$ Rydberg levels [see \eqref{eq:u0}] so that its value is negligible compared to $u_0$. 

We also emphasize that the eigenvalues of \eqref{eq:H_V} which correspond to the Born-Oppenheimer potential curves of the vdW interactions are isotropic, i.e do not depend on the angle $\theta$. However, in the presence of the magnetic-field along the axis $\bm{z}$, the Zeeman shift between the $m_j$ levels leads to anisotropic interactions.

In the limiting case $\theta=0$, as a consequence of the conservation of the total angular momentum, the doubly excited states $|\mone,\mone\rangle$, $|\pone,\pone\rangle$ are eigenstates of the vdW Hamiltonian, resulting in a typical blockade situation. In contrast, for $\theta=\frac{\pi}{2}$, the vdW Hamiltonian reduces to
\begin{equation}
H_V = \frac{1}{r^6}\left[\begin{matrix}u & 0 & 0 & c \\ 0 & u_0+c/3 &-c/3 & 0 \\  0 & -c/3 &u_0+c/3 &0 \\ c & 0 & 0 & u \end{matrix}\right],\label{eq:H_V_2}
\end{equation}
with $u\equiv u_0-c$. In the following, we consider this particular orientation as it corresponds to the simplest configuration with vdW couplings.

Our goal is now to study the dynamics of the system modeled by the Hamiltonian 
\begin{equation}
H\equiv \sum_{i=1,2} H_A^{(i)} + H_V\label{eq:H_tot}
\end{equation}
considering an initial state $|\Psi(t=0)\rangle = |g,g\rangle$. We note that the forces corresponding to vdW interactions are responsible for mechanical effects \cite{Glaetzle2012,Li2013}. Our model \eqref{eq:H_tot}, which neglects the motion of the atoms but also the spontaneous decay of the Rydberg level and black body transitions is therefore only valid in the \emph{frozen gas regime} \cite{Low2012} corresponding to short time scales, typically tens of micro seconds \cite{Beterov2009}. In the next subsection, we also comment on the possible influence of mechanical effects for our particular system.

For symmetry reasons, the dynamics is restricted to four states $|g,g\rangle$, \mbox{$|g,\mone\}\equiv\tfrac{1}{\sqrt{2}}(|g,\mone\rangle+|\mone, g\rangle)$},$|\mone,\mone\rangle$,$|\pone,\pone\rangle$ with the corresponding Hamiltonian 
\begin{equation}
H=\left[\begin{matrix}0 & \hbar\Omega/\sqrt{2} & 0 & 0\\
\hbar\Omega/\sqrt{2} & 0 & \hbar\Omega/\sqrt{2} & 0\\
0 & \hbar\Omega/\sqrt{2} & u/r^6 & c/r^6\\
0 & 0 & c/r^6 & u/r^6-2\hbar\Delta_{B}.
\end{matrix}\right].\label{eq:H_1}
\end{equation}
In order to estimate the influence of the coupling $c$ on the efficiency of the blockade mechanism, we are interested in the value of the average probability of double excitation
\begin{equation}
p_{2}(r,\theta)=\lim_{T\to\infty}\frac{1}{T}\int_0^T \left(\,|\langle\pone,\pone|\Psi(t)\rangle|^{2}+|\langle\mone,\mone|\Psi(t)\rangle|^{2}\right) dt.\label{eq:p2}
\end{equation}
In order to solve the dynamics and as presented in~\cite{Walker2008} in the absence of a magnetic-field, we now prediagonalize the Hamiltonian in the doubly-excited subspace:
\begin{equation}
H=\left[\begin{matrix}0 &\hbar\Omega/\sqrt{2} & 0 & 0\\
 \hbar\Omega/\sqrt{2} & 0 &\hbar\Omega_1(r)/2 &\hbar\Omega_2(r)/2\\
0 &\hbar\Omega_1(r)/2 & \lambda_1(r)& 0\\
0 &\hbar\Omega_2(r)/2 & 0 & \lambda_2(r),
\end{matrix}\right].\label{eq:H_2}
\end{equation}
written in the basis $|g,g\rangle$, $|g,\mone\}$,$|\lambda_1\rangle$,$|\lambda_2\rangle$ where the vdW eigenstates 
\begin{eqnarray}
|\lambda_1(r)\rangle&=& \cos\phi|\mone,\mone\rangle + \sin\phi|\pone,\pone\rangle,\label{eq:l1} \\
|\lambda_2(r)\rangle&=&\sin\phi|\mone,\mone\rangle - \cos\phi|\pone,\pone\rangle \label{eq:l2}, 
\end{eqnarray}
with $\tan[2\phi(r)]=c/(r^6\hbar\Delta_B)$ are directly coupled to the singly laser-excited state  $|g,\mone\}$ with space-dependent Rabi frequencies
\begin{eqnarray}
\Omega_1(r) &=& \sqrt{2}\Omega\cos\phi(r),\\
\Omega_2(r) &=& \sqrt{2}\Omega\sin\phi(r),
\end{eqnarray}
and have an energy
\begin{eqnarray}
\lambda_{1,2}(r) & = & \frac{u}{r^6}-\hbar\Delta_{B}\left(1\mp\sqrt{\left(\frac{c}{\hbar\Delta_Br^{6}}\right)^2+1}\right).\label{eq:lambdapm}
\end{eqnarray}

We illustrate the four level structure associated to the Hamiltonian \eqref{eq:H_2} in Fig.~\ref{fig:setup}(b). The advantage of this representation is that one can directly infer from the values of $\Omega_{1,2}$ and $\lambda_{1,2}$ whether the doubly excited state manifold can be populated. One can notice that in the limit of small distances $r\to0$,  the doubly excited states are energetically excluded from the dynamics whereas the state $|\mone,\mone\rangle$ can be resonantly excited at large distances $r\to\infty$ where the coupling $c/r^6$ becomes negligible.

This representation also illustrates the existence of a magic distance at $r=r_0$ at which the Zeeman shift compensates the vdW interactions resulting in a non-interacting vdW eigenstate $|\lambda_2\rangle$: $\lambda_2(r_0)=0$. Around this distance, the double excitation probability $p_2(r)$ will exhibit a resonance peak similar to the anti-blockade effect \cite{Ates2007} with the crucial difference that the non-interacting vdW eigenstate $|\lambda_2\rangle$ which is excited is a superposition of the two doubly excited states $|\mone,\mone\rangle$ and $|\pone,\pone\rangle$.

\subsection{Properties at the magic distance $r_0$}

We now describe the magic distance peak and show that it can be observed experimentally. First, its position, the magic distance $r_0$, corresponds to a zero of one vdW eigenstate: $\lambda_i(r_0)=0$ which leads to:
\begin{equation}
r_{0}=\left[\frac{u(1-\alpha)}{2\hbar\Delta_{B}}\right]^{1/6}\label{eq:r0}, 
\end{equation}
with $\alpha=c^2/u^2$. We note, that this solution only exists if 
\begin{equation}
\Delta_B u (1-\alpha)>0,\label{eq:condition}
\end{equation}
which illustrates the fact that the magic distance only exists when the magnetic-field competes with the vdW interactions ($\Delta_B$ and $B$ have opposite signs). In the following, we consider that the orientation of the magnetic-field $\sgn(B)$ is chosen to satisfy the condition \eqref{eq:condition} and derive analytical expressions describing the magic distance peak. From Eq.~\eqref{eq:r0}, one can notice that the magic distance $r_0$, depends on the value of the Zeeman shift $\Delta_B$: For small magnetic-fields, $r_0$ is larger than the blockade radius $r_b\equiv(u/\hbar\Omega)^{1/6}$ so that we expect the emergence of a peak in the non-blocked region. However, for large magnetic-fields, the magic distance leads to the formation of a peak in the supposedly blocked region, i.e. $r_0<r_b$.

Let us first describe the case $r_0<r_b$ as the behavior of the system is in total contradiction with the standard blockade criterion, i.e. $r<r_b\implies p_2(r)\to0$ \cite{Saffman2010}. In this regime, which corresponds to the situation shown in Fig.~\ref{fig:setup}(b), we can assume that around $r=r_0$ only the vdW eigenstate 
\begin{equation}
|\tilde{\lambda}\rangle= \begin{cases} |\lambda_2\rangle & \mbox{if } \alpha<1 \\ |\lambda_1\rangle & \mbox{if }  \alpha>1\end{cases},
\end{equation}
whose eigenvalue $\tilde{\lambda}$ vanishes at the magic distance: $\tilde{\lambda}(r_0)=0$, can be populated whereas the other vdW eigenstate is energetically excluded. Using the relation \mbox{$|\langle \tilde{\lambda}(r_0)|\mone,\mone\rangle|^2=\alpha/(1+\alpha)$} we then obtain the peak height
\begin{equation}
p_{2}(r_0)=\frac{3}{2}\alpha\frac{1+\alpha}{\left(1+2\alpha\right)^{2}}.\label{eq:p2max}
\end{equation}
The analytical expression Eq.~\eqref{eq:p2max} illustrates the key role of the vdW coupling $c$ in the magic distance phenomenon. Considering $P_{1/2}$ states, $\alpha$ is typically of the order of unity (see for example Fig.~\ref{fig:uc} for Rubidium atoms) so that the peak height $p_2(r_0)$ is significant. We also note that the value of $p_2(r_0)$ is independent of the magnetic-field.

We are now interested in the value of the peak width $\Delta r$ associated to the range of distances $(r)$ where $|\tilde{\lambda}\rangle$ can be populated. This quantity allows us to assess the importance of localization effects: In order to be observed, the peak width $\Delta r$ should be larger than the uncertainty  $\delta r$ in the distance $r$ between the atoms. We estimate the value of $\Delta r$ by considering that $|\tilde{\lambda}\rangle$ can be populated if its energy $\tilde{\lambda}$ is in the excitation bandwidth $[-\hbar\Omega,\hbar\Omega]$:  
\begin{equation}
\Delta r\sim\frac{2\hbar\Omega}{\tilde{\lambda}'(r_0)}\label{eq:Dr},
\end{equation}
with 
\begin{equation}
\tilde{\lambda}'(r_0)\equiv\left(\frac{\partial \tilde{\lambda}}{\partial r}\right)_{r=r_0}=\frac{-6u}{r_0^7}\left(\frac{1-\alpha}{1+\alpha}\right)\label{eq:slope}.
\end{equation}
The value of the derivative $\tilde{\lambda}'(r_0)$ allows therefore to estimate the possibility to observe the magic distance peak. Using Eqs.~\eqref{eq:r0} and.~\eqref{eq:slope}, we note that the width $\Delta r$ scales as $B^{-7/6}$: For very large magnetic-fields corresponding to $r_0\ll r_b$, the peak becomes narrow so that it might be difficult to observe it experimentally. However, in the limit of small magnetic-fields where $r_0\lesssim r_b$, $\Delta r$ can be larger than the uncertainty $\delta r$ and can be therefore experimentally resolved. We also note that, as the Rydberg atoms are not trapped, the vdW force $F=\lambda_1'(r_0)$ at the magic distance may induce mechanical effects~\cite{Li2013}. However, as in the case of the peak width, the strength of this force can be controlled via the value of the magnetic-field.

\begin{figure}
\centering{}\includegraphics[width=0.23\textwidth]{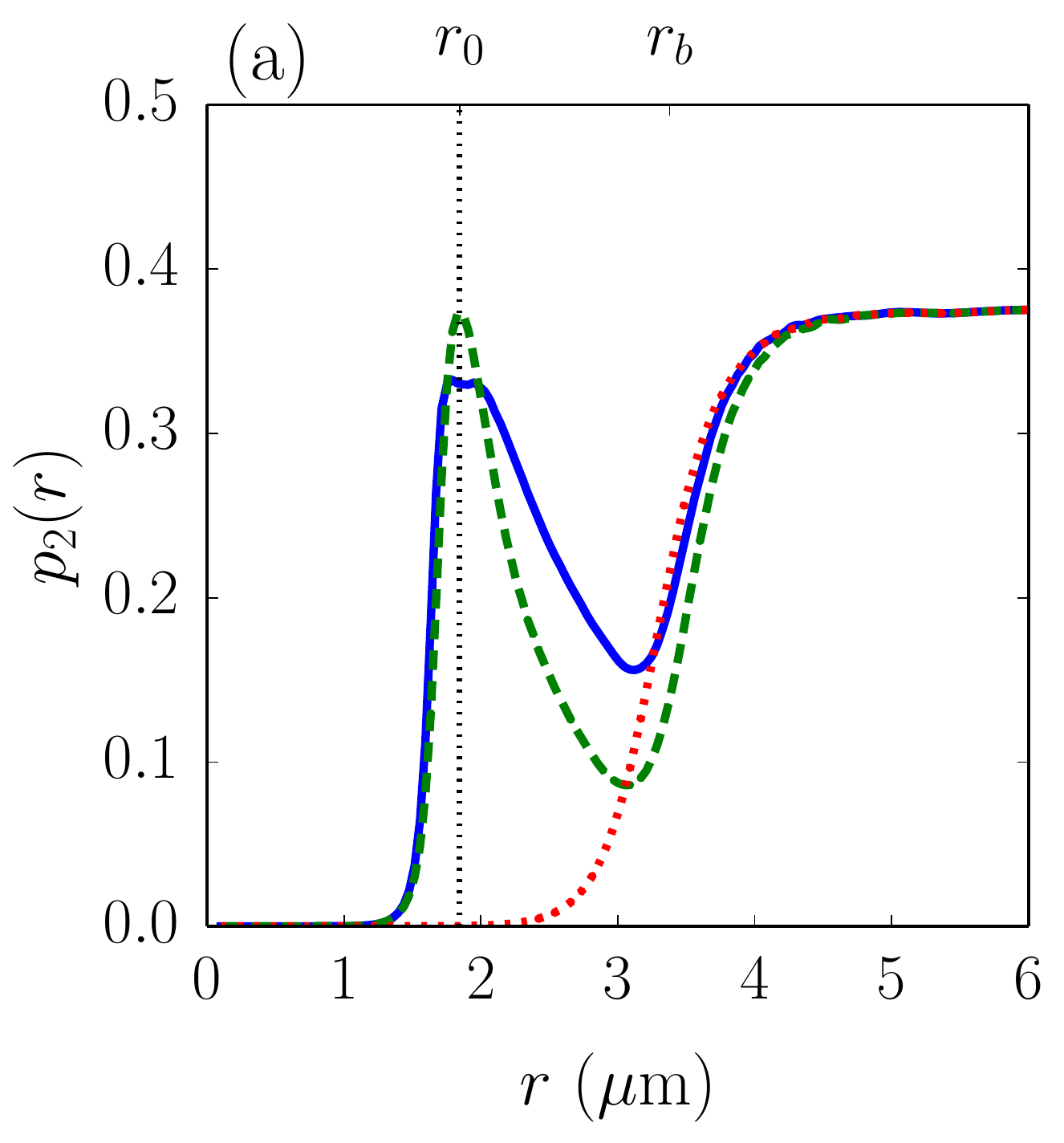}
\includegraphics[width=0.23\textwidth]{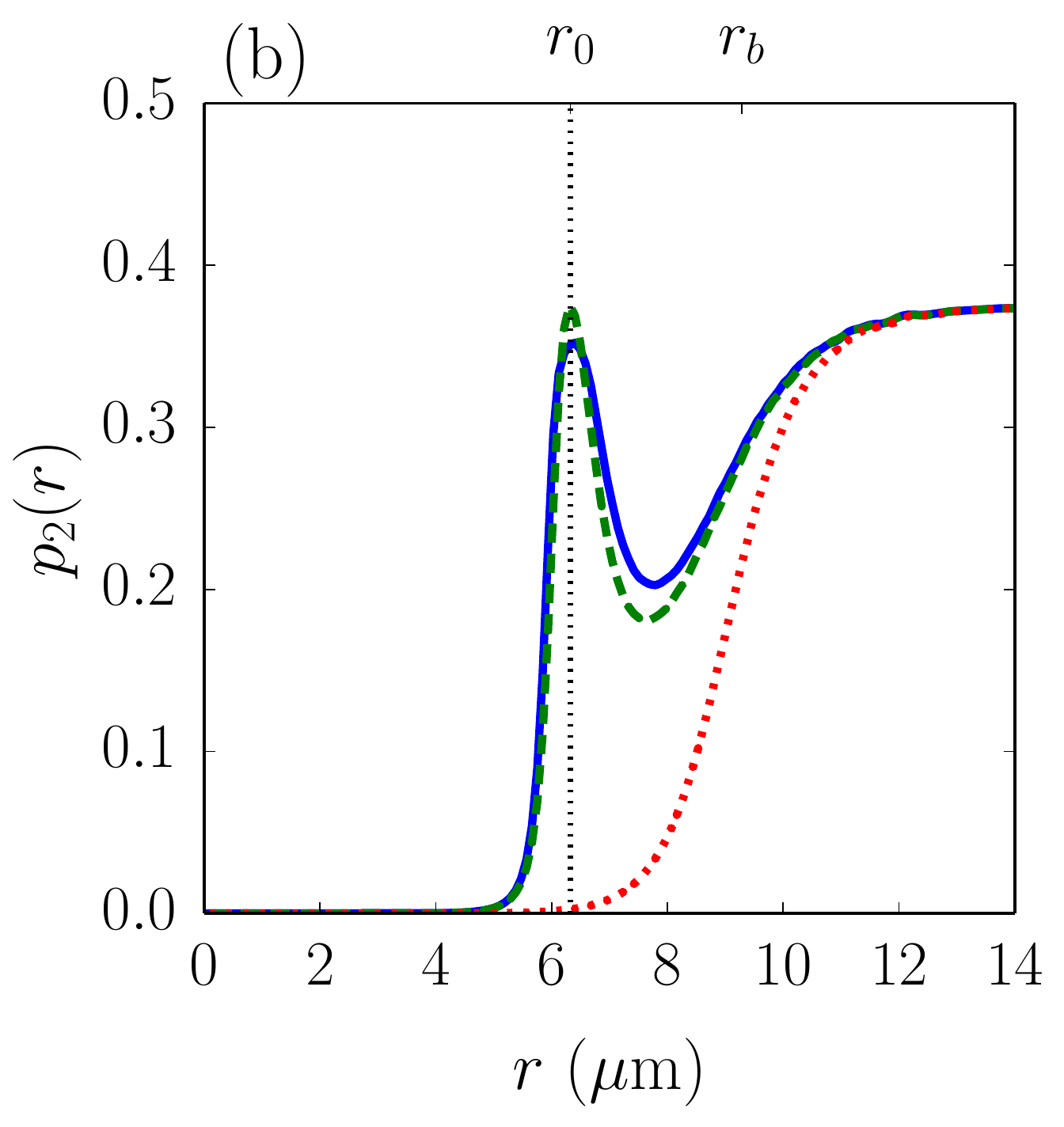}
\caption{(a) Mean double excitation probability $p_2(r)$ [Eq.~\eqref{eq:p2}] for $n=40$, $\Omega/(2\pi) = 0.2$ MHz, $B=0.2$
G. The blue solid line represents the full numerical result whereas the green dashed line is obtained from the effective Hamiltonian Eq.~\eqref{eq:H_eff}. Finally the red dotted line represents the usual blockade situation artificially obtained by setting $c=0$. (b) $p_2(r)$ for $n=70$ corresponding to $\alpha\sim 1.45$, a magnetic-field $B=-0.5$ G and same Rabi frequency.
\label{fig:p2}}
\end{figure}
In order to be more specific about the feasibility of a magic distance experiment, we now consider the example of Rubidium atoms. In this case, the diagonal element of the vdW interactions $u$ is always negative whereas $\alpha$ is smaller than $1$ for $n<43$ and larger otherwise [see Fig.~\ref{fig:uc}]. Consequently, the magic distance $r_0$ exists for $\Delta_B<0$ ($B>0$) if $n<43$ and  $\Delta_B<0$ ($B<0$) otherwise [condition \eqref{eq:condition}]. We represent in Fig.~\ref{fig:p2}(a) the probability of double excitation, obtained numerically, as a function of the distance $r$  for $n=40$, $\Omega/(2\pi) = 0.2$ MHz and $B=0.2$ G.The doubly excited states are blocked at small distances, and the limit $p_2\approx3/8$ obtained at large distances corresponds to the non-interacting limit. In contrast to a situation without the vdW coupling, obtained artificially by setting $c=0$, the value of $p_2$ is not monotonous as around the magic distance $r_0=1.85\ \mu$m [Eq.~\eqref{eq:r0}], a peak emerges illustrating the crucial impact of the vdW coupling $c$ on the Rydberg blockade. Its height $p_2(r_0)=0.3$ and width $\Delta r\sim0.7\,\mu$m are in good agreement with the analytical expressions Eqs.~\eqref{eq:p2max} and \eqref{eq:Dr}. Furthermore, the peak width $\Delta r$ is larger than the uncertainty $\delta r$ for typical experimental setups \cite{Viteau2011,Schauss2014,Ebert2014,Barredo2014,Hankin2014} whereas the characteristic length scale $\tilde{\delta} r\sim FT^2/(2\mu)\sim0.07 \mu$m  ($T=2\pi/\Omega$ is the typical time of the experiment and $\mu$ is the reduced mass of the atomic pair) corresponding to the vdW force is negligible compared to $\Delta r$. Consequently, for these parameters, the magic distance peak is accessible within state-of-the-art experiments. We note that, in this example, $\alpha=0.95$ is close to $1$ so that the derivative $\tilde{\lambda}'(r_0)$ associated to the width of the peak and the vdW force is strongly reduced compared to an anti-blockade configuration where $\tilde{\lambda}'(r_0)=-6u/r_0^7$, [see Eq.~\eqref{eq:slope}]. The possibility to observe the magic distance peak is however not restricted to a particular value of $\alpha$: Considering for example a larger value of the principal quantum number $n=70$ with $\alpha=1.5$, and $B=-0.5$ G [Fig.~\ref{fig:p2}(b)], we find $\Delta r\sim 1\ \mu\rm{m}\gg\delta r$ and $\tilde{\delta} r\sim0.05\ \mu \rm{m}\ll \Delta r$.

\begin{figure}
\centering{}\includegraphics[width=0.35\textwidth]{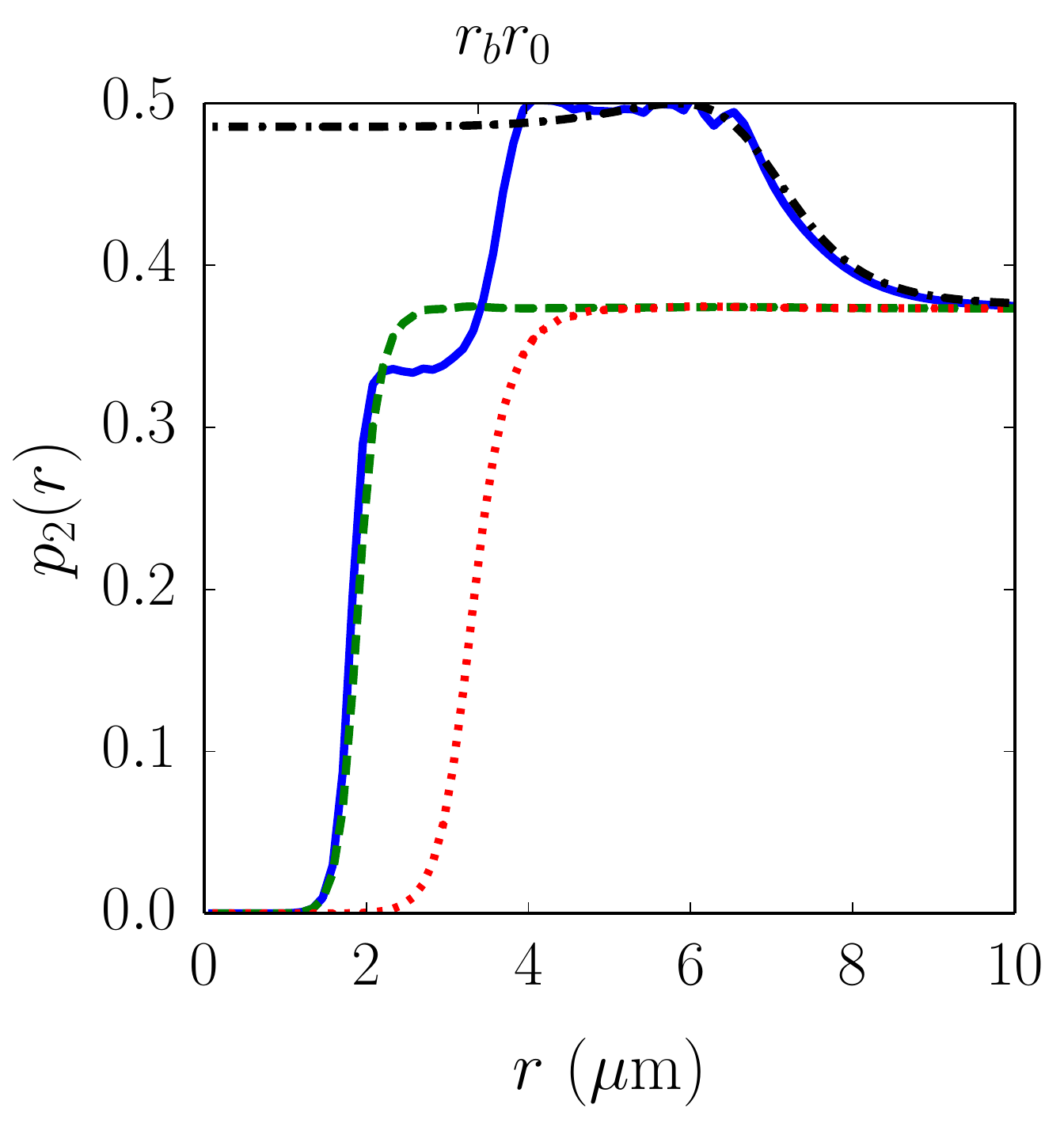}
\caption{Mean double excitation probability $p_2(r)$ [Eq.~\eqref{eq:p2}] for $n=40$, $\Omega/(2\pi) = 0.2$ MHz, $B=2$ mG with the same color code as in Fig.~\ref{fig:p2}. The peak appears at large distance $r>r_0,r_b$ and the analytical expression \eqref{eq:p2per} [black dash-dotted line] correctly describes its emergence from the non-interacting limit $r\to\infty$.\label{fig:p2b}}
\end{figure}

The expressions \eqref{eq:p2max}, \eqref{eq:Dr} and \eqref{eq:slope} were derived assuming that the peak emerges in the usually blocked region $r_0<r_b$. Let us now for the sake of completeness describe the limit of very small magnetic-fields leading to $r_0 > r_b$. Around the magic distance $r_0$,  $|\lambda_1\rangle$, $|\lambda_2\rangle$ can in this case be excited simultaneously. The resulting probability of double excitation, which can exceed the non-interaction limit $3/8$,   is obtained in the limit $\hbar\Omega\gg u/r^6,c/r^6,\hbar\Delta_B$ from first-order perturbation theory 
\begin{eqnarray}
p_{2}(r) & \approx & \frac{3}{8}+\frac{c^{2}}{\left(4\hbar\Delta_{B}r^6-u\right)^{2}+8c^{2}},\label{eq:p2per}
\end{eqnarray}
which shows in particular that $p_2(r)\le0.5$. An example of this situation is shown in Fig.~\ref{fig:p2b} for $n=40$ and $B=2$ mG. The doubly excited states are blocked at small distances, and the limit $p_2\approx3/8$ obtained at large distances corresponds to the non-interacting limit. The peak emerges in this situation around $r\approx4-6\,\mathrm{\mu m}$ and is well described by the analytical expression Eq.~\eqref{eq:p2per} (black dash-dotted line) for $r>r_0$. The width of this peak is in this example several microns, which shows that for a such small magnetic-field the effect of the vdW coupling on the dynamics is dominant.

\subsection{The effective potential as a blockade criterion}
The peak corresponding to a magic distance $r_0$ is due to the vdW coupling $c$ [see Eq.~\eqref{eq:p2max}] and is therefore a feature which goes beyond the usual Rydberg blockade. However, it is useful to define a generalized Rydberg blockade criterion that would allow one to determine directly from the values of the parameters whether the system is Rydberg blocked, replacing the usual criterion $r<r_b$. To do so and following the method of \cite{Walker2008} given in the absence of a magnetic-field, we consider the partially blocked regime $|\lambda_{1,2}|\gg \hbar \Omega$ and eliminate adiabatically the doubly excited states to obtain:
\begin{equation}
p_2(r) = \frac{\hbar^2\Omega^2}{2}\sum_{i=1,2}\frac{|\langle\mone,\mone|\lambda_{i}\rangle|^{2}}{\lambda_{i}^{2}}\label{p2adia}.
\end{equation}
This expression is indeed very instructive as the comparison with the value $\hbar^2\Omega^2/2V^2_\mathrm{eff}$ obtained from an effective usual Rydberg blockade Hamiltonian
\begin{equation}
H_\mathrm{eff}= \sum_{i=1,2} H_A^{(i)}+V_\mathrm{eff} |\mone,\mone\rangle\langle\mone,\mone|\label{eq:H_eff}
\end{equation}
 allows to define an effective potential
\begin{equation}
\frac{1}{V_{\mathrm{eff}}^{2}}\equiv\sum_{i=1,2}\frac{|\langle\mone,\mone|\lambda_{i}\rangle|^{2}}{\lambda_{i}^{2}}.\label{eq:V_eff}
\end{equation}
The effective potential summarizes the interplay between the vdW interactions and the magnetic-field and is particularly useful to know whether the system is blocked, the criterion $|V_\mathrm{eff}|>\hbar\Omega$ replacing the usual blockade criterion $r<r_b$. Moreover, even if its expression was derived in the partially blocked regime, the effective potential approach correctly describes the dynamics in the non-interacting limit $r\to\infty$ as in this case $V_\mathrm{eff}\to0$. Finally, this quantity was recently measured experimentally \cite{Barredo2014}, giving us a starting point to assess the relevance of our work, see next section. Finally, from Eq.~\eqref{eq:V_eff}, one can immediately notice that the zero of the effective potential indeed corresponds to the magic distance $r_0$. However, this approach cannot described properly the resulting peak $p_2(r_0)$. This is shown in Fig.~\ref{fig:p2}(a),(b) and \ref{fig:p2b} where the green dashed representing the effective approach allows to know whether the system is blocked or not but does not describe correctly the peak formed around $r_0$.

\section{Generalization to other angular momentum states\label{sec:dstates}}

The approach we used in the previous section  to study the blockade dynamics in the particular configuration $\theta=\pi/2$ for $P_{1/2}$ states can be generalized to any fine-structure manifold  ($S_{1/2}$, $P_{1/2}$, $P_{3/2}$, $D_{3/2}$, ...) and any angle $\theta$. In general, the vdW interactions within a given Zeeman manifold of Rydberg states can we written as (see App.\ref{app:vdW})
\begin{equation}
H_V(r,\theta)=\sum_{\alpha\beta\gamma\delta}V_{\alpha\beta;\gamma\delta}(r,\theta) |m_\alpha, m_\beta \rangle \langle m_\gamma,  m_\delta|,
\label{eq:Hvdw2}
\end{equation}
where we used the notation $|m_\alpha, m_\beta\rangle=|n^2L_j,m_{\alpha}\rangle_{z}\otimes|n^2L_j,m_{\beta}\rangle_{z}$. The matrix $V_{\alpha\beta;\gamma\delta}(r,\theta)$, which has dimension $(2j+1)^2$, accounts for the diagonal interactions and vdW couplings between the Zeeman sublevels. In the particular case of $\theta=0$ the matrix has block structure because of the conservation of the total angular momentum $M=m_\alpha+m_\beta=m_\gamma+m_\delta$. For arbitrary angle $\theta$ the total angular momentum $M$ can change by $0,\pm1,\pm2$ units.

\begin{figure}[tb]
\begin{centering}
\includegraphics[width=0.49\textwidth]{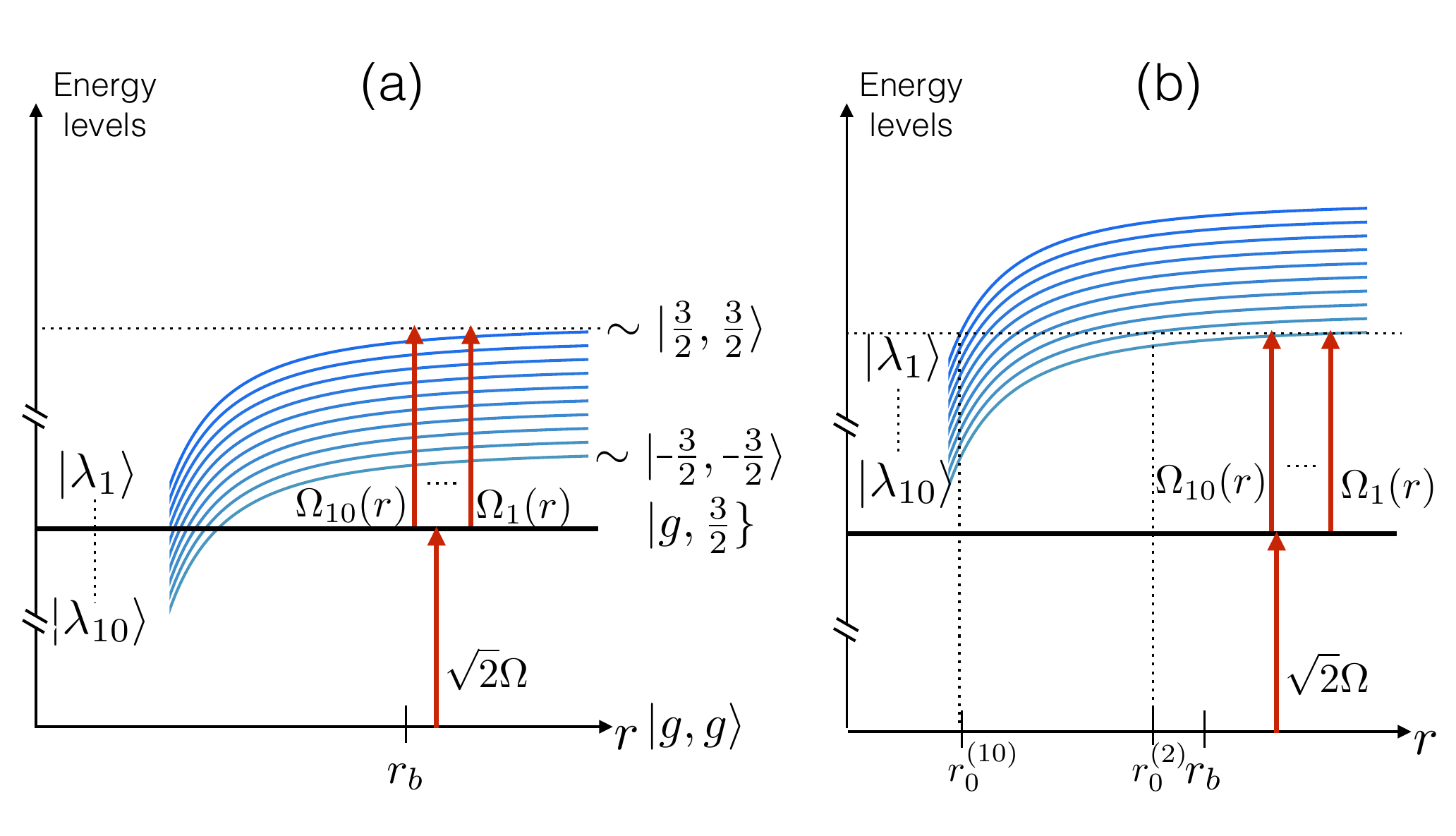}
\par\end{centering}
\caption{(a) Illustration of the blockade physics for laser-excited $P_{3/2}$ or $D_{3/2}$ Rydberg states with  (a) $|m_j=\pthree\rangle$ and (b) $|m_j=\mthree\rangle$ and a magnetic field pointing in the positive $z$ direction, i.e. $\bm{B}\cdot\bm{z}>0$. In panel (a) corresponding to the experimental setup \cite{Barredo2014}, the Zeeman splitting does not compete with the attractive vdW interactions so that there is no magic distance. In contrast, in panel  (b), a laser excitation to $|m_j=\mthree\rangle$ with the same magnetic field gives rise to a set a magic distances $r_0^{(i)}$.
\label{fig:Scheme_Palaiseau} }
\end{figure}

\begin{figure}
\includegraphics[width=0.45\textwidth]{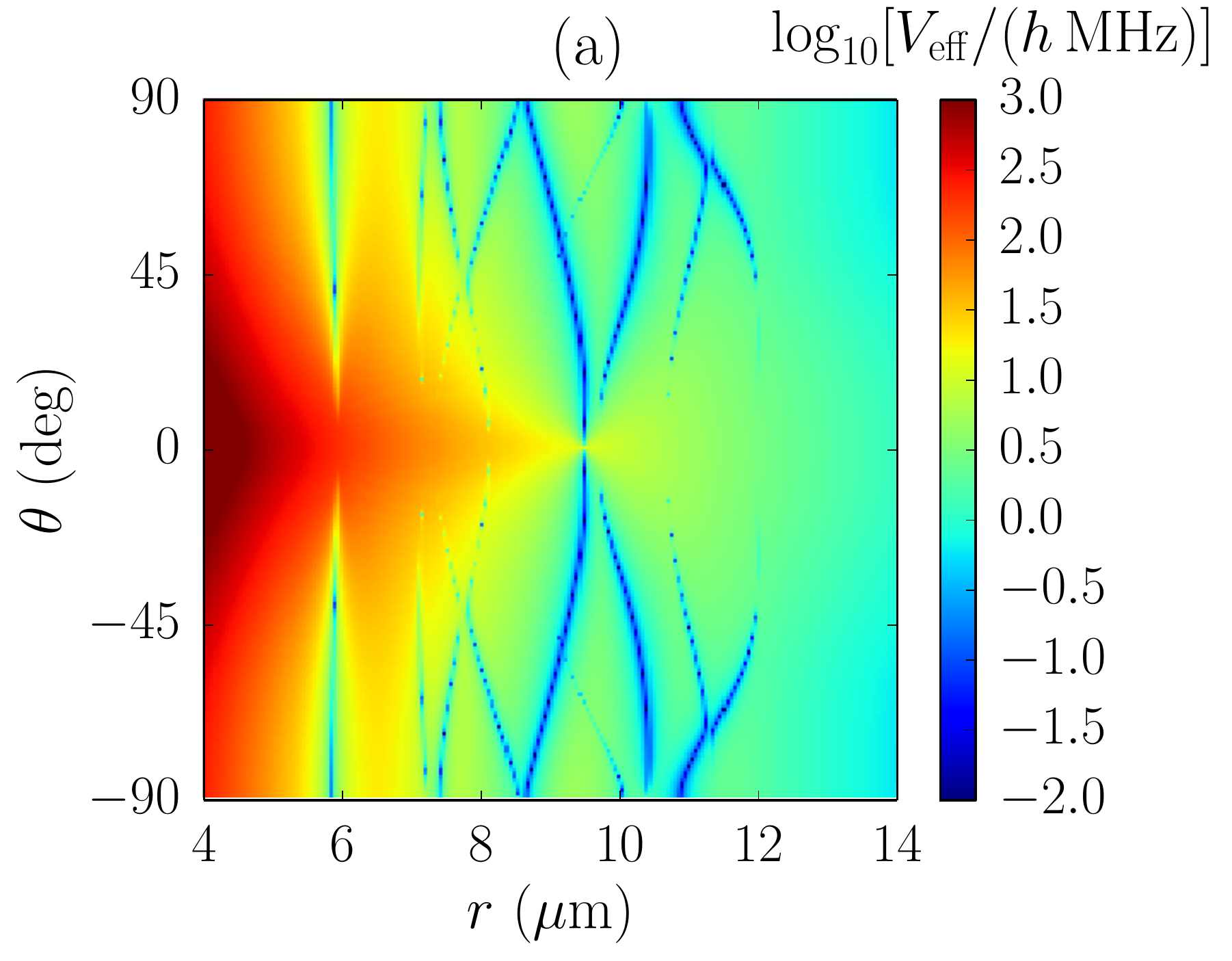}\vspace{0.5cm}
\includegraphics[width=0.45\textwidth]{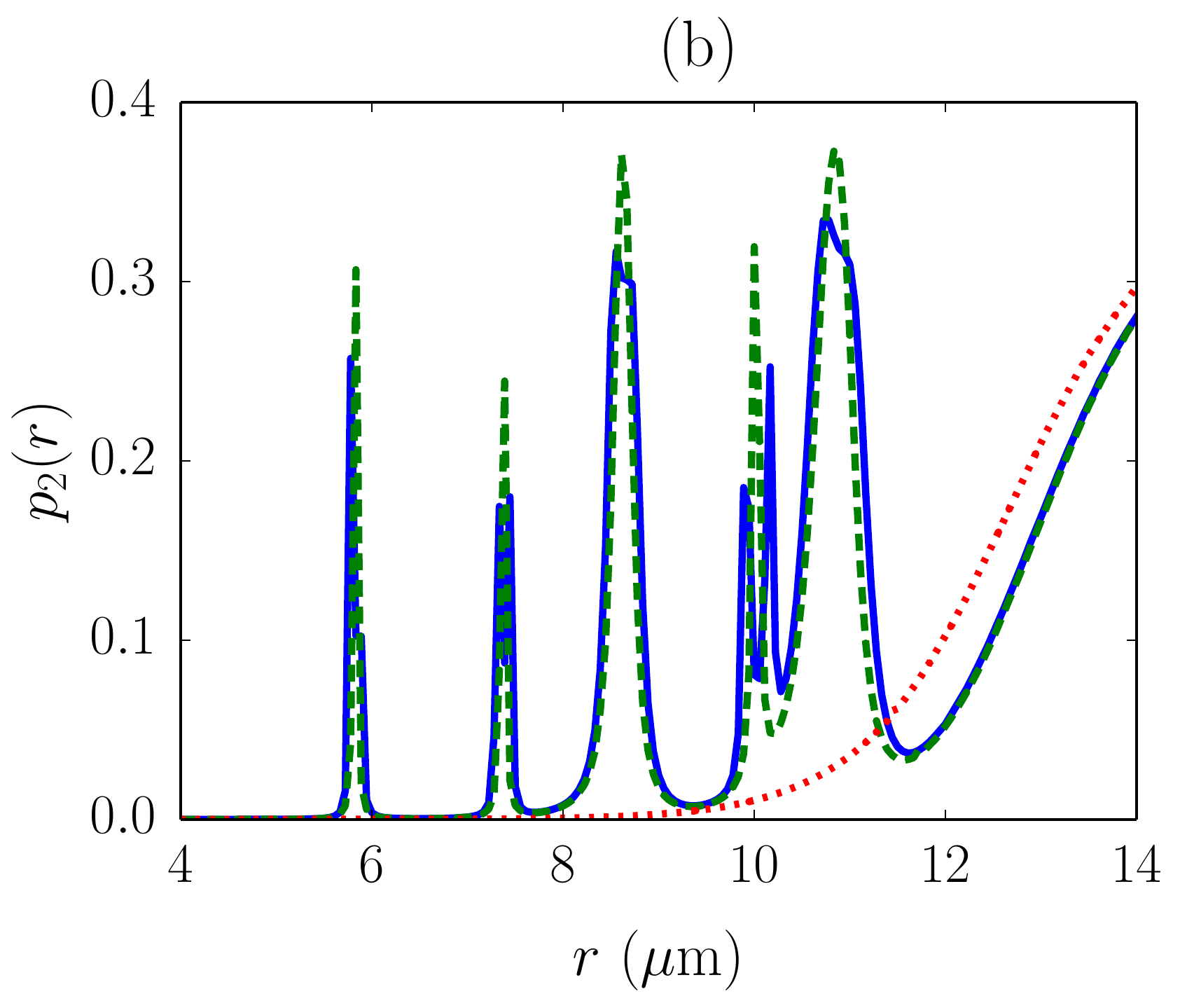}
\caption{(a) $V_{\mathrm{eff}}$ in MHz and logarithmic scale for $n=82$ and $B=3$ G for a laser-excitation to $|\mthree\rangle$. The positions of the magic distances correspond to the blue lines. (b) Corresponding probability of double excitation for $\theta=\pi/2$, illustrating the emergence of the peaks around the magic distances. The graphical conventions are defined in Fig.~\ref{fig:p2}\label{fig:Vvsrtheta}. }
\end{figure}

In the remainder of this section we investigate $P_{3/2}$ and $D_{3/2}$ Rydberg manifolds with $j=3/2$ and interpret a recent experiment performed in Palaiseau~\cite{Barredo2014}  demonstrating an anisotropic Rydberg blockade for $D_{3/2}$ states.  In the case of $j=3/2$ Rydberg states, the vdW Hamiltonian $H_V$ can, in principle, be written as a $16\times16$ matrix (see App.~\ref{app:vdW}). However, as the anti-symmetric states of the type $1/\sqrt{2}(|m_1,m_2\rangle-|m_2,m_1\rangle$) with $m_1\neq m_2$ are not coupled to the laser-excited state $|\pthree,\pthree\rangle$, we can reduce the basis to $10$ states. Therefore, the two vdW eigenstates $|\lambda_{1}\rangle$, $|\lambda_{2}\rangle$ for $P_{1/2}$ states of the previous section are simply replaced by a set of vdW eigenstates $\{|\lambda_i\rangle\}_{i=1}^{10}$ of the vdW Hamiltonian of Eq.~\eqref{eq:Hvdw2}. For a particular magnetic field direction and a given angle $\theta$, these 10 eigenstates will lead to a set of magic distances $\{\bm{r}_i\}$, illustrated in Fig.~\ref{fig:Scheme_Palaiseau}, and consequently to a multi-peak structure of the Rydberg excitation probability (see Fig.~\ref{fig:Vvsrtheta}). These eigenstates are then used to derive an effective potential which can predict the behavior of the system apart from its zeros, which are the magic distances.

As an illustration, let us now analyze how one can understand the recent experimental demonstration of anisotropic Rydberg blockade~\cite{Barredo2014} which were obtained using a laser excitation to the Rydberg state \mbox{$|82^2D_{3/2},m_j=\pthree\rangle\equiv|\pthree\rangle$}, with two atoms separated by a distance $r=12\ \mu$m and an arbitrary angle $\theta$ with respect to the magnetic-field  $\bm{B}=B\bm{z}$ with $B=3$ G. Following the method introduced in the previous section, we diagonalize the total Hamiltonian $H$ in the subspace formed by the doubly excited states and obtain $10$ vdW eigenstates ${|\lambda_i\rangle}$ with Rabi frequencies $\Omega_{i}=\sqrt{2}\Omega\langle\lambda_{i}|\pthree,\pthree\rangle$. The corresponding graphical representation,  shown in Fig.~\ref{fig:Scheme_Palaiseau} (a) illustrates that as the interactions are attractive (see Appendix~\ref{app:vdW}) and the Zeeman shift between the laser-excited $|\pthree,\pthree\rangle$ and the other doubly excited states is negative, this configuration does not allow for magic distances.
Consequently, the competition between the vdW interactions and the magnetic-field can be described by an effective potential:
\begin{equation}
\frac{1}{V_{\mathrm{eff}}^{2}}=\sum_{i=1}^{10}\frac{|\langle\pthree,\pthree|\lambda_{i}\rangle|^{2}}{\lambda_{i}^{2}}.
\end{equation}
We note that by following this approach we neglect the excitation of the singly excited states which can only be excited from the pair states (for example the excitation of $|g,\pone\rangle$ from  $|\pthree,\pone\rangle$). However, we checked numerically that their effect is negligible in this situation.

\begin{figure}
\centering{}\includegraphics[width=0.85\columnwidth]{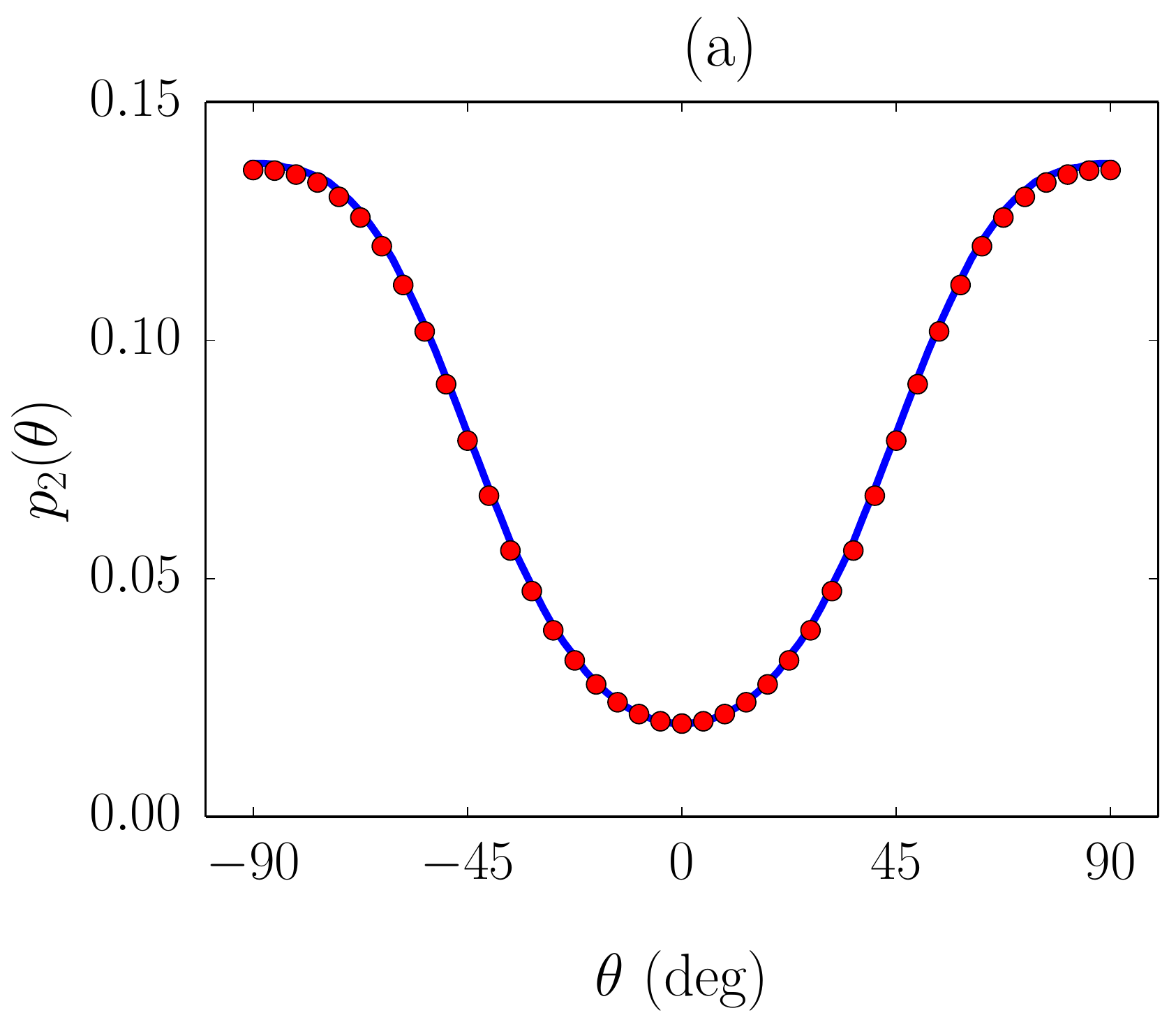}\\
\includegraphics[width=0.85\columnwidth]{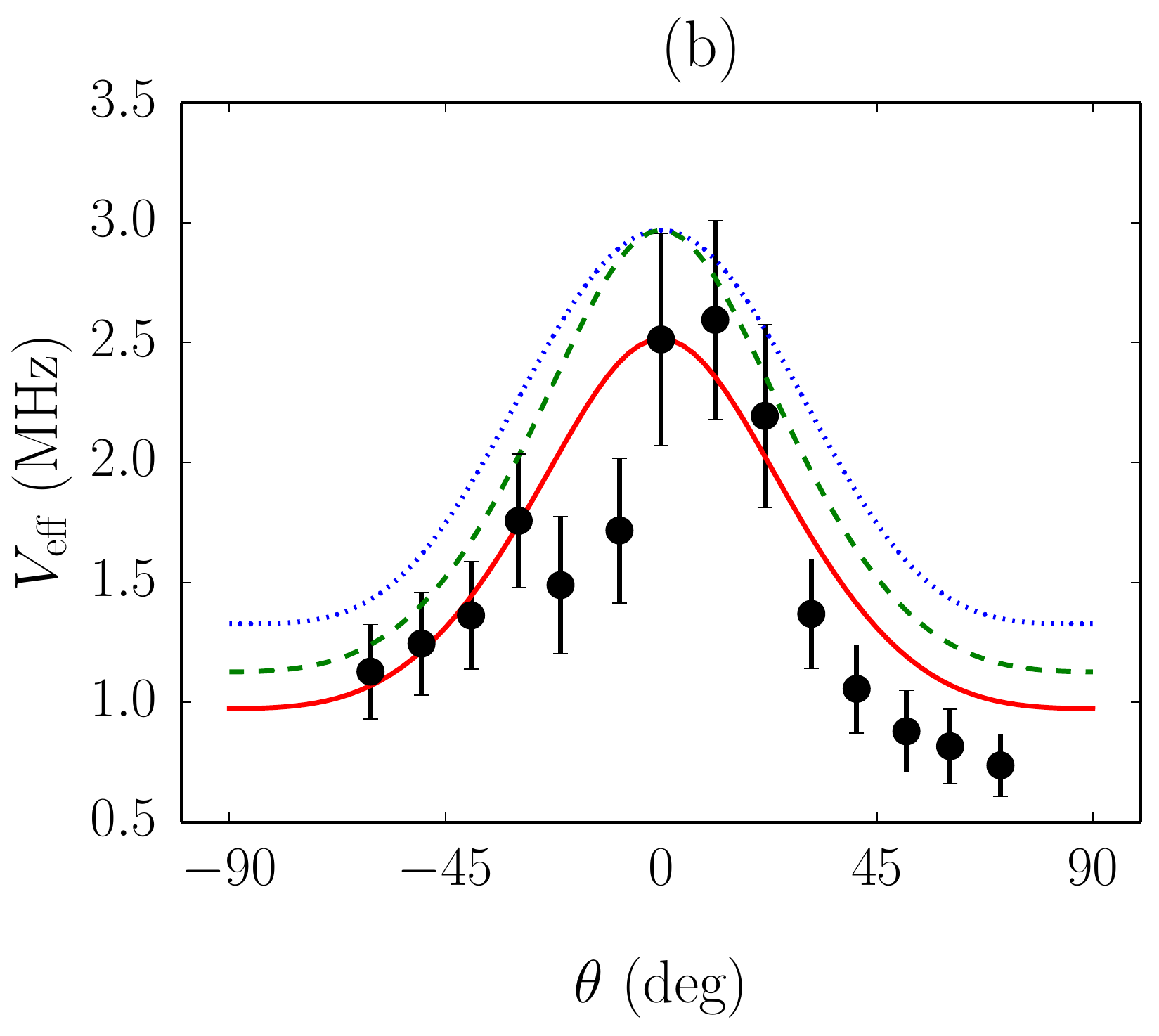}
\caption{(a) Average pair excitation probability obtained by the diagonalization
of the full Hamiltonian (blue solid line) and by the effective potential (red circles).
(b) $V_{rr}$ (blue dotted line), $V_{\mathrm{eff}}$ (green dashed lines) , $V_{\mathrm{eff}}$
(red solid line) which relies on the experimental determination of $C_{6}$ and experimental
data \cite{Barredo2014} (dark circles)\label{fig:palaiseau}.}
\end{figure}

We now compare in Fig.~\ref{fig:palaiseau}(a), the probability of double excitation $p_2(\theta)$ obtained by the effective potential approach and by the full Hamiltonian, for the experimental value of the Rabi frequency $\Omega/(2\pi)= 0.8$ MHz. The system is in this case in the partially blocked regime $p_2<0.2$ and the result obtained with the effective potential is in good agreement with the exact result. This allows us to compare the theoretical value of the effective potential to the corresponding experimental quantity obtained by fitting the dynamics with an effective Hamiltonian \cite{Barredo2014}. The result is shown in Fig.~\ref{fig:palaiseau}(b): The blue dotted line corresponds to the naive diagonal element \mbox{$V_{rr}\equiv\langle\pthree,\pthree|H_V|\pthree,\pthree\rangle$} of the vdW Hamiltonian and clearly disagrees with the experimental results. The green dashed line represents the effective potential calculated from the theoretical value of the vdW coefficients, defined in App.~\ref{app:vdW}. 
Finally, we obtained the red solid line using an experimental correction for these coefficients which is obtained for $\theta=0$ as there is no vdW coupling in this case (the small discrepancy between experimental and theoretical values at $\theta=0$ was already noticed in~\cite{Beguin2013} and attributed to mechanical effects~\cite{Li2013}). The green curve captures the qualitative influence of the mixing between Rydberg states whereas the red curve gives a quantitative agreement. These results confirm the relevance of our approach in a situation where the influence of the vdW coupling is not very significant as illustrated by the small difference between the naive and the effective potential in Fig.~\ref{fig:palaiseau}(b).

Let us now investigate the case of a laser-excitation to the Rydberg state $|\mthree\rangle$, represented schematically in Fig.~\ref{fig:Scheme_Palaiseau}(b). In this situation, the magnetic-field competes with the attractive vdW  interactions leading to the existence of a set of magic distances. The corresponding effective potential is shown in Fig.~\ref{fig:Vvsrtheta}(a) as a function of $r$ and $\theta$, where its zeros correspond to the positions of the magic distances as a function of $\theta$. Except for $\theta=0$ where there is no vdW coupling, this situation leads to the existence of up to $9$ magic distances. We represent in panel (b), the probability $p_2(r)$ of double excitation for $\theta=\frac{\pi}{2}$ showing the emergence of a peak around each of these magic distances. Furthermore, in order to minimize localization and mechanical effects, the position of the magic distances and the width of the corresponding peaks can be increased by applying a smaller magnetic-field (this behavior can be understood graphically with Fig.\ref{fig:superpair}(b), the width of the peaks being related to the slope of the energy levels where they cross the zero-energy line, see also Sec.~ \ref{sec:method}).

\section{Many-body effects at the magic distance\label{sub:magic}}
Now that we have characterized the behavior of the system at the magic distance for two atoms and have shown in particular that the resulting peak can be observed with Rydberg $p$, $d$ states, and in many geometric configurations, let us investigate the many-body scenario. In the absence of vdW coupling $c$, the Rydberg blockade mechanism is responsible for the excitation of a super-atom, which corresponds to a superposition of singly excited states with a collective enhancement of the light-atom coupling \cite{Lukin2001,Low2012}. We now show that at the magic distance, the many-body dynamics results in the collective excitation of a superposition of pair states, generalizing the concept of super-atom.

\begin{figure}
\begin{centering}
\includegraphics[height=4cm]{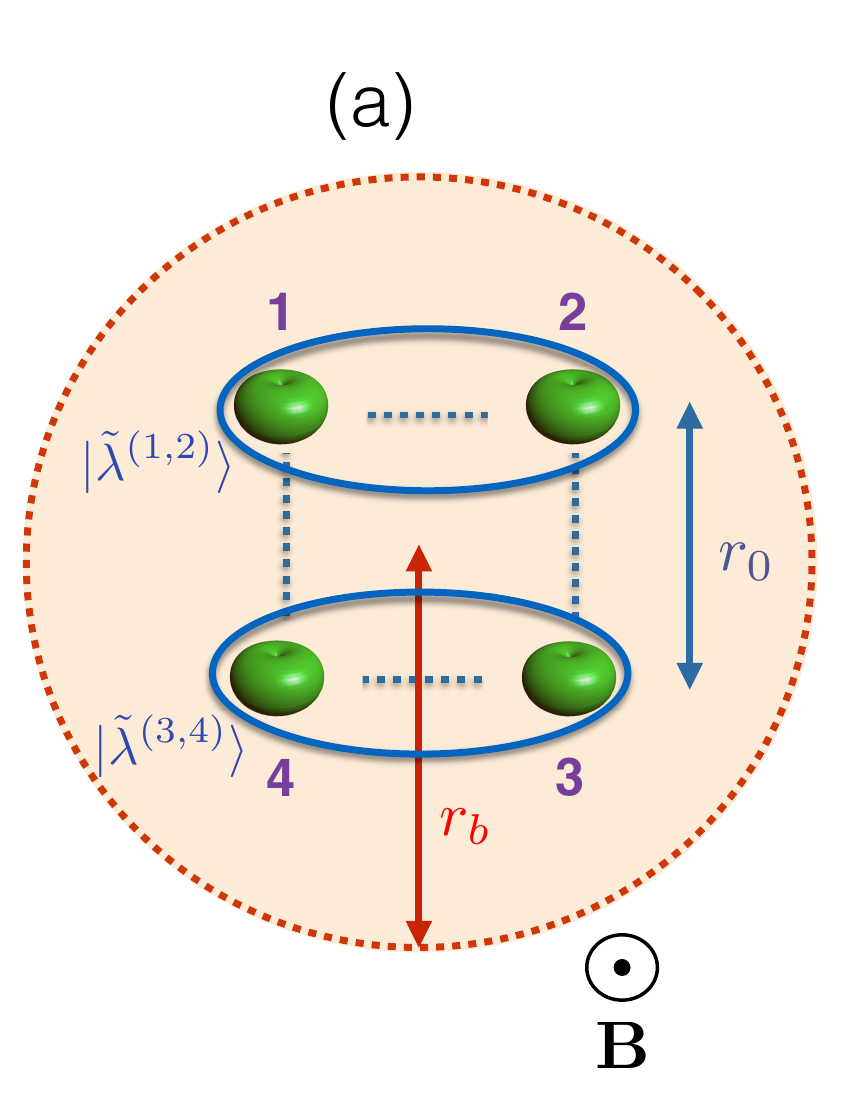}
\includegraphics[height=4cm]{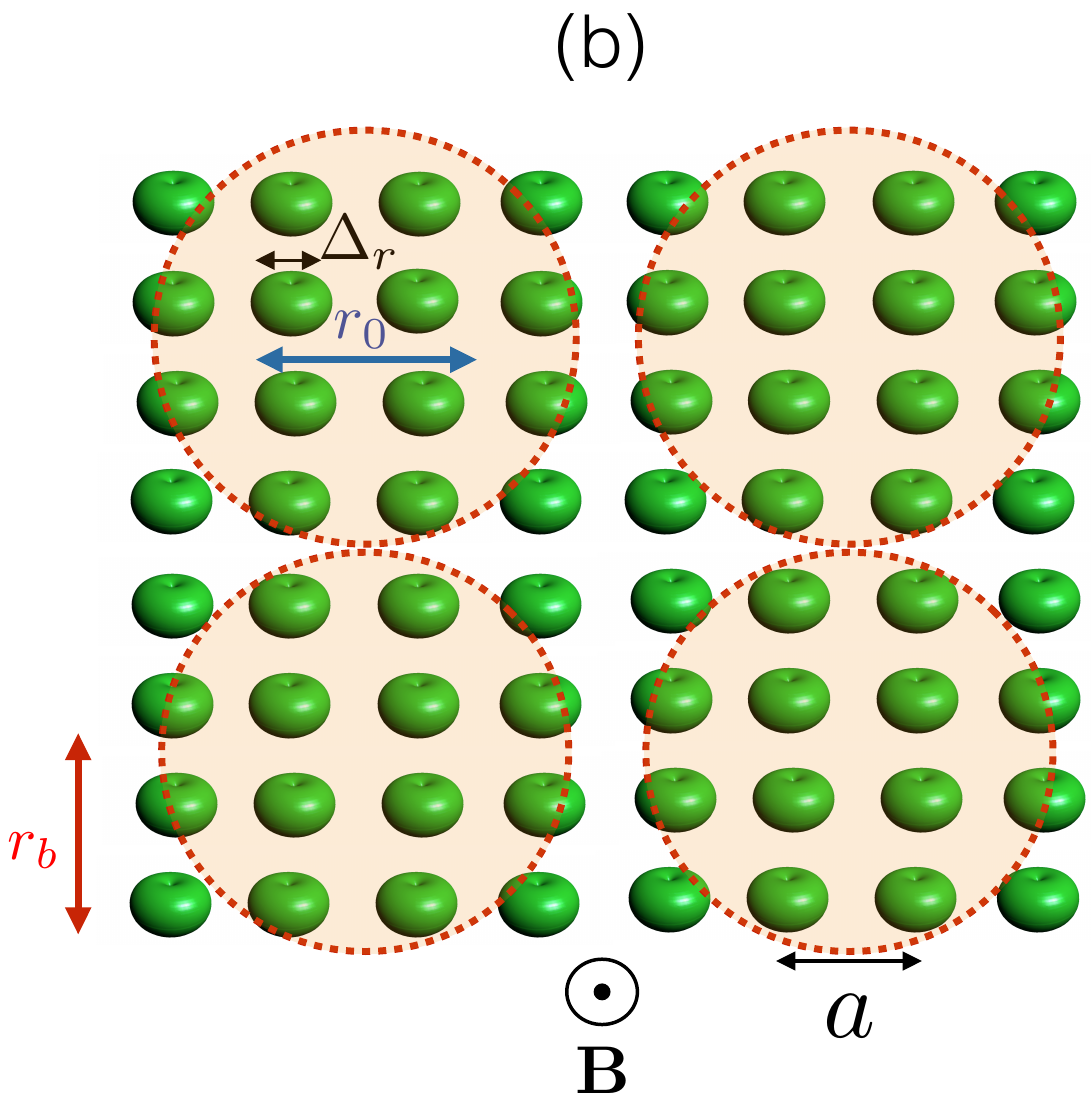}
\includegraphics[height=6cm]{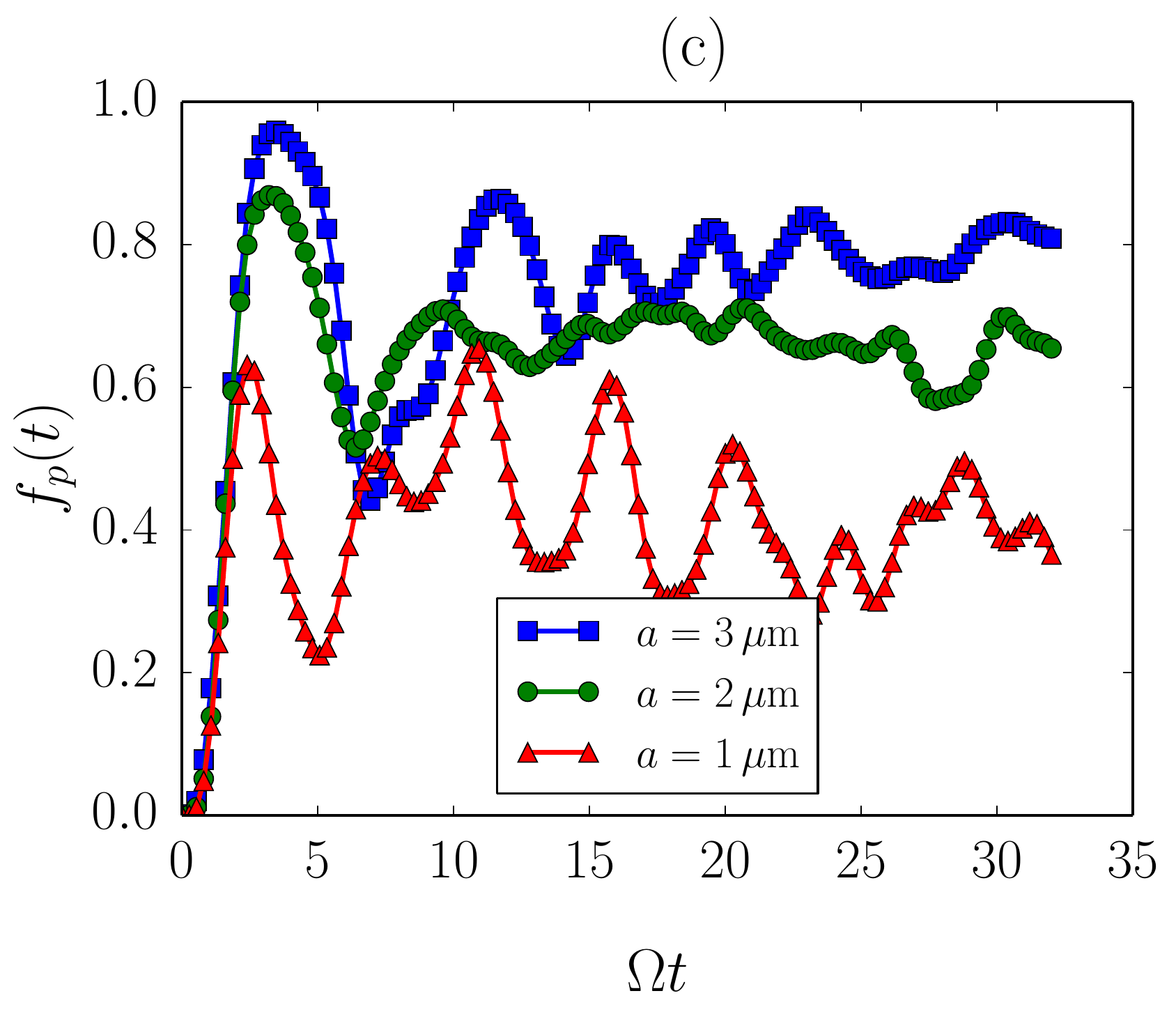}
\par\end{centering}
\caption{(a) Four atoms placed on a square of side $r_0$ cannot be excited simultaneously due to the non-additive properties of the vdW Hamiltonian. This blockade mechanism is responsible for the excitation of a superposition of pair states. (b) For a larger number of atoms, we can divide the system in an ensemble of independent blockade spheres in order to obtain the three state model. (c) Super pair oscillation $f_p(t)$  in the lattice regime $a>\Delta_r$ for $N=6$ and Rubidium $40P_{1/2}$ states with $\Omega=2\pi\times0.2$MHz and three different lattice spacings $a=1,2,3\,\mu$m and $r_0(B)=a$. \label{fig:superpair}}
\end{figure}

In order to understand the many-body dynamics associated to the magic distance phenomenon, it is instructive to first consider 4 atoms ($i=1,2,3,4$) placed at the corners of a square of side $r_0$ [see Fig.~\ref{fig:superpair}(a)] and laser-excited to a $P_{1/2}$ Rydberg state $|\mone\rangle$. The direction of the magnetic-field is perpendicular to the plane formed by the square so that for all pairs of atoms, the angle $\theta$ is fixed to $\pi/2$ whereas the value of the magnetic-field is adjusted to fulfill the condition $r_0<r_b$.  According to the results of Sec.~\ref{sec:method}, for each pair of atoms $(i,j)$ separated by $r_0$, there is a non-interacting vdW eigenstate $|\tilde{\lambda}^{(i,j)}\rangle$.  However, the double pair states such as $|\tilde{\lambda}^{(1,2)}\rangle\otimes|\tilde{\lambda}^{(3,4)}\rangle$ (shown schematically in Fig.~\ref{fig:superpair}(a)) are not vdW eigenstates of the total interaction Hamiltonian and are therefore energetically excluded from the dynamics \footnote{This blockade effect is due to the fact that the vdW eigenstates cannot be written as a product state in the uncoupled basis [Eq.~\eqref{eq:l1}]}. Consequently,  considering that initially all atoms are in the ground-state $|g\rangle$ and given the symmetries of the problem, the dynamics is restricted to three states: The initial state $|\tilde{\psi}_{g}\rangle=|g\ldots g\rangle$, the familiar super-atom state $|\tilde{\psi}_{s}\rangle  =  \tfrac{1}{2}\sum_i|g\ldots\mone^{(i)}\ldots g\rangle$  and the super pair state $|\tilde{\psi}_{p}\rangle  =  \tfrac{1}{2}\sum_{|\bm{r}_i-\bm{r}_j|=r_0}|g\ldots\tilde{\lambda}^{(i,j)}\ldots g\rangle$.

The above discussion can be easily extended to the case of an arbitrary large number of atoms $N$, which is shown schematically in Fig.~\ref{fig:superpair}(b). First, as in the super-atom model \cite{Low2012}, we can describe the many-body dynamics dividing the system into an ensemble of blockade spheres of radius $r_b$, which evolve independently. Second, as shown in the case of four atoms, inside each blockade sphere two non-interacting vdW states cannot be excited simultaneously and the dynamics is restricted to three symmetric states: 
\begin{eqnarray}
|\tilde{\psi}_{g}\rangle & = & |g\ldots g\rangle\\
|\tilde{\psi}_{s}\rangle & = & \frac{1}{\sqrt{N_b}}\sum_{i}|g\ldots\mone^{(i)}\ldots\rangle\\
|\tilde{\psi}_{p}\rangle & = & \frac{1}{\sqrt{M_b}}\sum_{\ ||\bm{r}_i-\bm{r}_j|-r_0|<\Delta_r}|g\ldots\lambda^{(i,j)}_{1}\ldots g\rangle\label{eq:psip}
\end{eqnarray}
where $|\tilde{\psi}_{g}\rangle$ is the initial state, $|\tilde{\psi}_{s}\rangle$ is the super-atom state with $N_b$ the number of atoms in the blockade sphere and $|\tilde{\psi}_{p}\rangle$
is the ``super pair state'' which is the coherent superposition
of the $M_b$ non-interacting vdW eigenstates formed at the magic distance inside the blockade sphere.

The many-body Hamiltonian associated to our three-state model reduces to 
\begin{eqnarray}
H&=&\frac{\hbar\Omega}{2}\left[\begin{matrix}0 & \sqrt{N_b} & 0\\\sqrt{N_b} & 0 & \epsilon\\0 & \epsilon & 0\end{matrix}\right]_{|\tilde{\psi}_{g}\rangle,|\tilde{\psi}_{s}\rangle,|\tilde{\psi}_{p}\rangle}\label{eq:H_super},
\end{eqnarray}
with $\epsilon^2=4 M_b\alpha/[N_b(1+\alpha)]$. This Hamiltonian describes the collective enhancement of the light-atom coupling by the vdW interactions. Considering $|\Psi(t=0)\rangle=|\tilde{\psi}_{g}\rangle$, we can write the wave function at any time $t$:
\begin{equation}
|\Psi(t)\rangle=\left[\begin{matrix}\frac{\Omega^{2}}{\Omega_{s}^{2}} \left(N_b \cos {\frac{\Omega_{s} t}{2}} + \epsilon^{2}\right)\\- \frac{i \Omega}{\Omega_{s}} \sqrt{N_b} \sin {\frac{\Omega_{s} t}{2}}\\\frac{\sqrt{N_b} \epsilon}{\Omega_{s}^{2}} \Omega^{2} \left(\cos {\frac{\Omega_{s} t}{2}} - 1\right)\end{matrix}\right]_{|\tilde{\psi}_{g}\rangle,|\tilde{\psi}_{s}\rangle,|\tilde{\psi}_{p}\rangle},
\label{eq:phi_superpair}
\end{equation}
with the enhanced Rabi frequency:
\begin{equation}
\Omega_{s}=\Omega\sqrt{N_b+\epsilon^2}.\label{eq:Omegas}
\end{equation}
In the absence of vdW coupling ($\alpha=0$), the value of the coupling $\epsilon$ to the super pair state $|\tilde{\psi}_p\rangle$ vanishes and we obtain the familiar super-atom limit. However, with $P_{1/2}$ states, $\alpha$ is of the order of unity so that the existence of the magic distance is responsible for the excitation of the super pair state. We emphasize that the strength of the coupling $\epsilon$ to the super pair state depends on the number of magic distances $M_b$ per blockade sphere and thus on the geometric arrangement of the atoms. In a ``lattice configuration'' where the width $\Delta r$ of the magic distance peak is smaller than the lattice spacing $a$ [Fig.\ref{fig:superpair}(b)], $\epsilon$ can takes significant value when the magic distance $r_0$ is commensurate with the lattice and is negligible otherwise. In the limit where the width $\Delta r$ is larger than the typical inter-atomic distance $a$, the value of $\epsilon$, which can be obtained in the continuum limit, becomes proportional to $\Delta r$ and can be easily controlled via the magnetic-field.

Our three-state model shows the main effect of the magic distance at the many-body level, which is a collective excitation of the super pair state directly related to the geometric arrangement of the atoms.  However, this model only gives a first approximation of the dynamics because it neglects the influence of finite-size effects \footnote{For example the symmetric expression of the super pair state \eqref{eq:psip} assumes that there is no boundary effect.}. In order to estimate these effects, we now study numerically the dynamics of a small system of $6$ atoms placed on a 1D chain where the value of the magnetic-field is adjusted to satisfy: $r_0(B)=a>\Delta_r$. Fig.~\ref{fig:superpair}(b) shows the time-evolution of the super pair fraction $f_p$ which is defined here as the probability to have two neighboring sites simultaneously excited. As predicted by our model, the super pair fraction oscillates with a frequency and a mean value which depends on the lattice spacing $a$, the progressive attenuation of the oscillations being due to finite-size effects. As the value of the lattice spacing $a$ decreases, we note that the frequency of the oscillations increases while the corresponding amplitude is reduced. This is in agreement with the prediction of our model [Eq.~\eqref{eq:psip}] as the number of atoms $N_b$ per blockade sphere scales as $a^{-1}$ whereas the geometric factor $\epsilon^2\propto M_b/N_b$ does not vary significantly when the value of $a$ changes.

We have therefore successfully generalized the concept of super-atom in presence of a magic distance. Due to the presence of a non interacting vdW eigenstate, the doubly excited states can be populated with an enhanced Rabi frequency $\Omega_s$ which depends on the number of atoms $N_b$ per blockade sphere but also on the spatial distribution of the atoms via the geometric factor $\epsilon^2$.

\section{Conclusion}
In summary, in this paper we have described the effect of the vdW couplings with $p$ and $d$ Rydberg states and generalized the fruitful concept of Rydberg blockade. We have discussed the existence and have analyzed the properties of magic distances, and their effect on many body quantum dynamics. In particular, our three-state model introduces the key features of the magic distance at the many body level and points to a more detailed investigation regarding its consequences on the quantum simulation of many-body systems, such as, for example, the crystallization of Rydberg excitations in optical lattices ~\cite{Schauss2012,Schauss2014} or the statistical properties of these excitations in atomic clouds~\cite{Malossi2014,Schempp2014}.

We thank T. Lahaye, A. Browaeys and J. Zeiher for interesting discussions. Work at Innsbruck is supported
by the ERC-Synergy Grant UQUAM and SFB FOQUS of the Austrian Science Fund. B. V. acknowledges the Marie Curie
Initial Training Network COHERENCE for financial support.
\appendix

\section{Anisotropic van der Waals Hamiltonians\label{app:vdW}}
\subsection{General considerations}

We present here the properties of the vdW interactions~\cite{Walker2008} and diagonalize the corresponding Hamiltonian to generalize the results of \cite{Singer2005}. We focus on distances comparable to the usual blockade Radius for which interactions are in the MHz
regime, that is to say much smaller than the fine-structure splitting.  We also do not consider the case of a F\"orster resonance ~\cite{Saffman2010}: as a consequence, interactions are ruled by vdW interactions and the atoms are characterized by the same first three quantum
numbers $n,\ell,j$. Within these two assumptions, we can treat the dipole-dipole operator in second-order perturbation theory and obtain the interaction Hamiltonian in the fine-structure basis $(|m_\alpha, m_\beta\rangle$)
\begin{equation}
H_V(r,\theta)=\sum_{\alpha\beta\gamma\delta}V_{\alpha\beta;\gamma\delta}(r,\theta) |m_\alpha ,m_\beta \rangle \langle m_\gamma , m_\delta|,
\end{equation}
with 
\begin{equation}
V_{\alpha\beta;\gamma\delta}(r,\theta)=\frac{1}{r^{6}}\sum_{i}C_{6}^{(i)}\langle m_\alpha ,m_\beta |\mathcal{D}_{i} | m_\gamma, m_\delta \rangle\label{eq:Hvdw}.
\end{equation}
The channel $i$ refers to the second and third quantum number ($\ell_s,j_s$) and ($\ell_t,j_t$) of the intermediary states, the corresponding vdW coefficient $C_{6}^{(i)}$ being:
\[
C_{6}^{(i)}=\sum_{n_s,n_t}\frac{e^{4}}{-\delta_{\alpha,\beta}}(R_{nlj}^{n_s\ell_sj_s}R_{n\ell j}^{n_t\ell_tj_t})^{2}
\]
where the radial matrix elements $R$ are calculated using model potentials
\cite{Marinescu1994} combined with spectroscopy measurements \cite{Li2003,Han2006}
and $\delta_{\alpha,\beta}$ is the F\"orster defect \cite{Saffman2010}. The matrices $\mathcal{D}_{i}$
describe the angular dependent part: Its $(2j+1)^{2}$ matrix elements depend explicitly on $\theta$, the
angle between the quantization and the molecular axis:
\begin{eqnarray}
\mathcal{D}_{i}&=&\sum_{m_s,m_t}\mathcal{M}_{i}|m_s,m_t\rangle\langle m_s,m_t|\mathcal{M}_i^\dag\\
&+& \delta_{(\ell_s,j_s)\neq(\ell_t,j_t)}\sum_{m_s,m_t}\mathcal{M}_{i}|m_t,m_s\rangle\langle m_t,m_s|\mathcal{M}_i^\dag
\end{eqnarray}
where $\mathcal{M}_{i}$ describes the angular momentum dependence
of the dipole-dipole operator:
\begin{equation}
\begin{split}\langle m_{\alpha},m_{\beta}|\mathcal{M}_{i}|m_s,m_t\rangle= & -\sqrt{\frac{24\pi}{5}}(2\ell+1)^2(2j+1)^2\\
\times & (-1)^{s-m_{\alpha}}\sqrt{(2\ell_s+1)(2j_{s}+1)}\\
\times & (-1)^{s-m_{\beta}}\sqrt{(2\ell_{t}+1)(2j_t+1)}\\
\times & \ws{\ell,\ell_s,1,j_s,j,s}\wt{\ell_s,1,\ell,0,0,0}\\
\times & \ws{\ell,\ell_t,1,j_t,j,s}\wt{\ell_t,1,\ell,0,0,0}\\
\times & \sum_{\mu,\nu}\wt{j_t,1,j,m_t,\nu,-m_\beta}\wt{j_s,1,j,m_s,\mu,-m_\alpha}\\
 & C_{\mu,\nu;\mu+\nu}^{1,1;2}Y_{2}^{\mu+\nu}(\vartheta,\varphi)^{*}.
\end{split}
\end{equation}

\subsection{$j=1/2$}

In the case of the $P_{1/2}$ and $S_{1/2}$ states, the vdW matrix
is written in the basis $|\mone,\mone\rangle$, $|\mone,\pone\rangle$, $|\pone,\mone\rangle$,
$|\pone,\pone\rangle$. The intermediary states are given in Table \ref{tab:Channels12}
\begin{table}[h]
\begin{ruledtabular}
\begin{tabular}{ccc}
channel $i$ & $l_sj_s\ l_tj_t$ for $S_{1/2}$ states & $l_sj_s\ l_tj_t$ for $P_{1/2}$ states\tabularnewline
1 & $P_{1/2}P_{1/2}$ & $P_{1/2}S_{1/2}$ \tabularnewline
2 & $P_{1/2}P_{3/2}$ & $P_{1/2}D_{3/2}$\tabularnewline
3 & $P_{3/2}P_{3/2}$ & $D_{3/2}D_{3/2}$ \tabularnewline
\end{tabular}\end{ruledtabular}\caption{Channels involved in the case of $S_{1/2}$ and $P_{1/2}$ Rydberg
pair states. \label{tab:Channels12}}
\end{table}
whereas the dimensionless matrices $\mathcal{D}_{i}$ can be written in the form:
\begin{small} 
\begin{eqnarray*}
\mathcal{D}_{1} & = &\frac{1}{27}\left[\begin{matrix}- \cos {2 \theta} + \frac{7}{3} & \sin {2 \theta} & \sin {2 \theta} & - 2 \sin^{2} {\theta}\\\sin {2 \theta} & \cos {2 \theta} + \frac{5}{3} & \cos {2 \theta} + \frac{5}{3} & - \sin {2 \theta}\\\sin {2 \theta} & \cos {2 \theta} + \frac{5}{3} & \cos {2 \theta} + \frac{5}{3} & - \sin {2 \theta}\\- 2 \sin^{2} {\theta} & - \sin {2 \theta} & - \sin {2 \theta} & - \cos {2 \theta} + \frac{7}{3}\end{matrix}\right]
\\
\mathcal{D}_{2} & = &\frac{2}{27}\left[\begin{matrix}- 2 \sin^{2} {\theta} + \frac{14}{3} & - \sin {2 \theta} & - \sin {2 \theta} & 2 \sin^{2} {\theta}\\- \sin {2 \theta} & 2 \sin^{2} {\theta} + \frac{10}{3} & 2 \sin^{2} {\theta} - \frac{8}{3} & \sin {2 \theta}\\- \sin {2 \theta} & 2 \sin^{2} {\theta} - \frac{8}{3} & 2 \sin^{2} {\theta} + \frac{10}{3} & \sin {2 \theta}\\2 \sin^{2} {\theta} & \sin {2 \theta} & \sin {2 \theta} & - 2 \sin^{2} {\theta} + \frac{14}{3}\end{matrix}\right]\\
\mathcal{D}_{3} & = & \frac{1}{27} \left[\begin{matrix}- \cos {2 \theta} + \frac{25}{3} & \sin {2 \theta} & \sin {2 \theta} & - 2 \sin^{2} {\theta}\\\sin {2 \theta} & \cos {2 \theta} + \frac{23}{3} & \cos {2 \theta} + \frac{5}{3} & - \sin {2 \theta}\\\sin {2 \theta} & \cos {2 \theta} + \frac{5}{3} & \cos {2 \theta} + \frac{23}{3} & - \sin {2 \theta}\\- 2 \sin^{2} {\theta} & - \sin {2 \theta} & - \sin {2 \theta} & - \cos {2 \theta} + \frac{25}{3}\end{matrix}\right].
\end{eqnarray*}
\end{small}
leading to Eq.~\eqref{eq:H_V} with
\begin{eqnarray}
u_0&=& \frac{4 C^{\left(1\right)}_{6}}{81} + \frac{28 C^{\left(2\right)}_{6}}{81} + \frac{22 C^{\left(3\right)}_{6}}{81}\label{eq:u0}\\
c &= & - \frac{2 C^{\left(1\right)}_{6}}{27} + \frac{4 C^{\left(2\right)}_{6}}{27} - \frac{2 C^{\left(3\right)}_{6}}{27}\label{eq:c}
\end{eqnarray}
The values of $u_0$, $u=u_0-c$ and $c$ used in the main text are shown in Fig.~\ref{fig:uc} for Rubidium $P_{1/2}$ states. In contrast to $S_{1/2}$ states where the vdW coupling is negligible $|c|\ll|u|$, the three terms are of the same order of magnitude.

\begin{figure}
\centering{}
\includegraphics[width=0.35\textwidth]{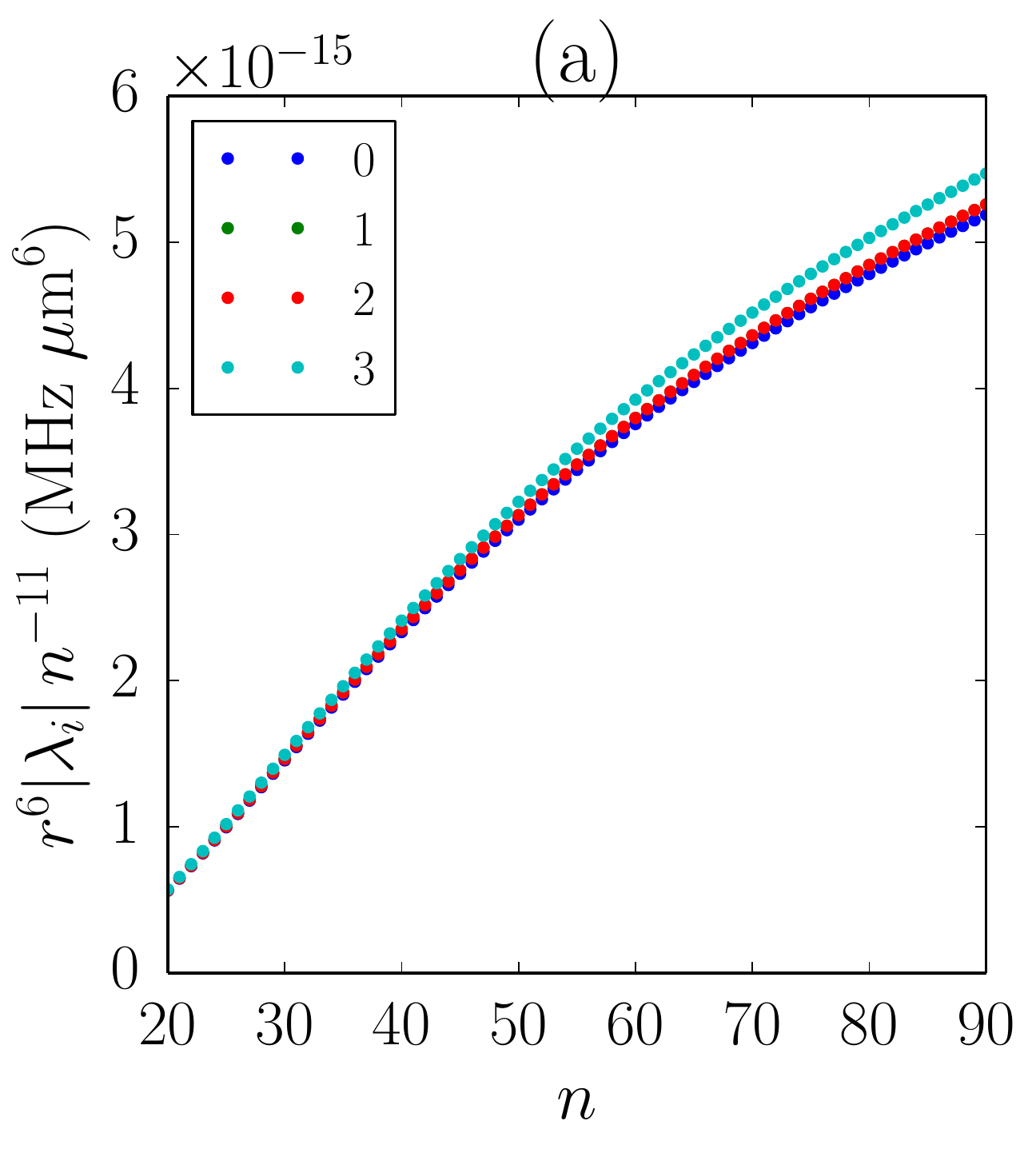}\\
\includegraphics[width=0.35\textwidth]{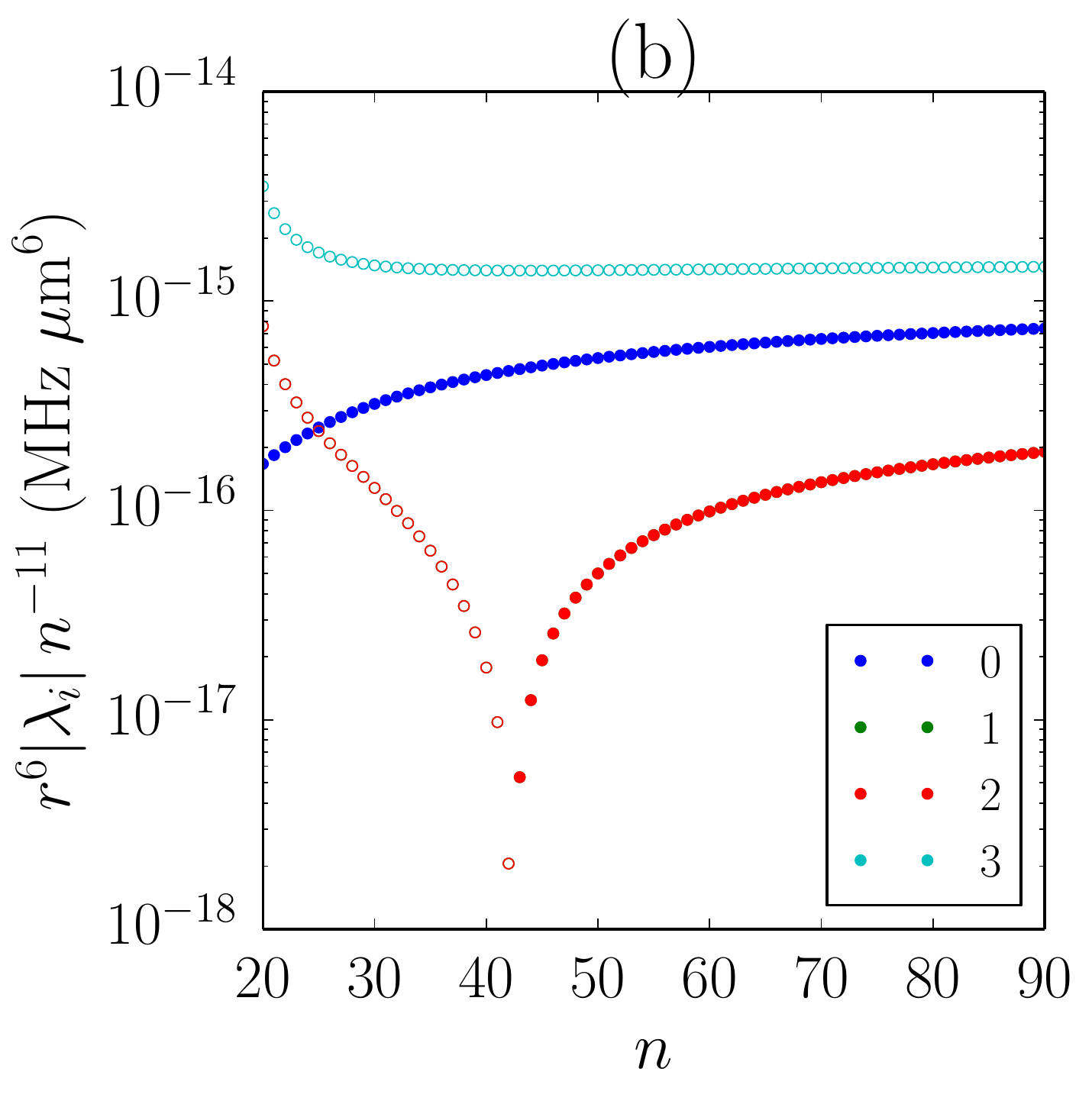}
\caption{Energies of the vdW eigenstates for Rubidium $S_{1/2}$ (a) and $P_{1/2}$ (b) states as a function of the principal quantum number $n$\label{fig:Energies_p}. Empty (full) symbols represent negative (positive) values.}
\end{figure}

It is also particularly instructive to diagonalize the vdW Hamiltonian to know for example whether interactions are attractive or repulsive. The easiest way to do it is to consider the basis where the quantization axis aligned
with the directions of the atoms ($\theta=0$). We can get then the
expression of the eigenstates for an arbitrary $\theta$ with an appropriate change of
basis. The eigenvalues are 
\begin{eqnarray}
\lambda_{0} & = & \frac{1}{r^{6}}\left(\frac{4C_{6}^{(2)}}{9}+\frac{2C_{6}^{(3)}}{9}\right)\\
\lambda_{1} & = & \lambda_{2}=\frac{1}{r^{6}}\left(\frac{4C_{6}^{(1)}}{81}+\frac{28C_{6}^{(2)}}{81}+\frac{22C_{6}^{(3)}}{81}\right)\\
\lambda_{3} & = & \frac{1}{r^{6}}\left(\frac{16C_{6}^{(1)}}{81}+\frac{4C_{6}^{(2)}}{81}+\frac{34C_{6}^{(3)}}{81}\right).
\end{eqnarray}
with the corresponding eigenstates 

\begin{eqnarray}
|\lambda_{0}\rangle & = & \frac{1}{\sqrt{2}}\left(|\pone,\mone\rangle-|\mone,\pone\rangle\right)\\
|\lambda_{1}\rangle & = & |\mone,\mone\rangle\\
|\lambda_{2}\rangle & = & |\pone,\pone\rangle\\
|\lambda_{3}\rangle & = & \frac{1}{\sqrt{2}}\left(|\pone,\mone\rangle+|\mone,\pone\rangle\right).
\end{eqnarray}
The state $|\lambda_{0}\rangle$ could be seen as a ``F\"orster zero''  \cite{Walker2005} as the first channel does not contribute to its energy. $\lambda_{0}$ is indeed of the same order of magnitude as
the other energies as shown in Fig.~\ref{fig:Energies_p} representing
the distance-independent term $r^{6}\lambda_{i}$ for $S_{1/2}$ state
(a) and $P_{1/2}$ states (b), as a function of the principal quantum
number $n$. For $S_{1/2}$ states, the energies are almost degenerate
whereas  $P_{1/2}$ states represent a promising option to study effects which go beyond
the usual blockade picture as the eigenenergies are different and
do not even have the same sign for $n>42$.

\subsection{$j=3/2$}
In the case of $P_{3/2}$ and $D_{3/2}$ states, six
channels participate to the vdW interactions. They are shown in Table \ref{tab:Channels-p32}.
\begin{table}[h]
\begin{ruledtabular}
\begin{tabular}{cccccc}
channel $i$ &  & $l_sj_s\ l_tj_t$ for $P_{3/2}$ states & $l_sj_s\ l_tj_t$ for $D_{3/2}$ states &  & \tabularnewline
1 &  & $S_{1/2}S_{1/2}$  & $P_{1/2}P_{1/2}$  &  & \tabularnewline
2 &  & $S_{1/2}D_{3/2}$  & $P_{1/2}P_{3/2}$  &  & \tabularnewline
3 &  & $S_{1/2}D_{5/2}$  & $P_{1/2}F_{5/2}$  &  & \tabularnewline
4 &  & $D_{3/2}D_{3/2}$  & $P_{3/2}P_{3/2}$  &  & \tabularnewline
5 &  & $D_{3/2}D_{5/2}$  & $P_{3/2}F_{5/2}$  &  & \tabularnewline
6 &  & $D_{5/2}D_{5/2}$  & $F_{5/2}F_{5/2}$  &  & \tabularnewline
\end{tabular}\end{ruledtabular}
\caption{Channels involved in the case of $P_{3/2}$ and $D_{3/2}$ states.
\label{tab:Channels-p32}}
\end{table}
We do not write the $16\times16$ $\mathcal{D}_{i}$ matrices (which are identical for $P_{3/2}$ and $D_{3/2}$ states) as they can be
obtained easily from standard symbolic routines.
 However, let us diagonalize the total vdW Hamiltonian
in the particular case $\theta=0$ where it is composed
of block corresponding to each value of total angular momentum $M=m_{\alpha}+m_{\beta}$.
The symmetry of the problem with respect to the transformation $|m_{\alpha},m_{\beta}\rangle\to|\minus m_{\alpha},\minus m_{\beta}\rangle$
allows us to restrict the study to the case $M\ge0$. Its means in
particular that the eigenvalues associated to $M>0$ are degenerate. As
for $j=1/2$ states , the eigenstates for arbitrary $\theta$ are
obtained by the appropriate change of basis.

\subsection{$M=3$}

For $M=3$, there is only one state 
\begin{equation}
|\lambda_{0}\rangle=|\pthree,\pthree\rangle
\end{equation}
with interaction energy (we omit the pre-factor
$\frac{1}{r^{6}})$ 
\begin{equation}
\lambda_{0}=\frac{4C_{6}^{(3)}}{15}+\frac{4C_{6}^{(4)}}{625}+\frac{136C_{6}^{(5)}}{1875}+\frac{84C_{6}^{(6)}}{625}.
\end{equation}

\subsection{$M=2$}

In this case, the two eigenstates are simply 
\begin{eqnarray}
|\lambda_{1}\rangle & = & \frac{1}{\sqrt{2}}\left[|\pone,\pthree\rangle+|\pthree,\pone\rangle\right]\\
|\lambda_{2}\rangle & = & \frac{1}{\sqrt{2}}\left[|\pone,\pthree\rangle-|\pthree,\pone\rangle\right]
\end{eqnarray}
with eigenvalues 
\begin{eqnarray}
\lambda_{1} & = & \frac{2C_{6}^{(2)}}{225}+\frac{12C_{6}^{(3)}}{25}+\frac{16C_{6}^{(4)}}{5625}+\frac{42C_{6}^{(5)}}{625}+\frac{84C_{6}^{(6)}}{625}\\
\lambda_{2} & = & \frac{2C_{6}^{(2)}}{25}+\frac{4C_{6}^{(3)}}{75}+\frac{22C_{6}^{(5)}}{375}+\frac{36C_{6}^{(6)}}{125}.
\end{eqnarray}
\subsection{$M=1$}

In this case, three interacting states are coupled
$|\pthree,\mone\rangle,|\mone,\pthree\rangle$
and $|\pone,\pone\rangle$. The eigenstates are 
\begin{eqnarray*}
|\lambda_{3}\rangle & = & \frac{\cos\alpha}{\sqrt{2}}\left[|\pthree,\mone\rangle+|\mone,\pthree\rangle\right]+\sin\alpha|\pone,\pone\rangle\\
|\lambda_{4}\rangle & = & \frac{-\sin\alpha}{\sqrt{2}}\left[|\pthree,\mone\rangle+|\mone,\pthree\rangle\right]+\cos\alpha|\pone,\pone\rangle\\
|\lambda_{5}\rangle & = & \frac{1}{\sqrt{2}}\left[|\pthree,\mone\rangle-|\mone,\pthree\rangle\right]
\end{eqnarray*}
with 
\begin{eqnarray*}
\cos(2\alpha) & = & \frac{\delta_{\alpha}}{\sqrt{\delta_{\alpha}^{2}+C_{\alpha}^{2}}}\\
\sin(2\alpha) & = & \frac{C_{\alpha}}{\sqrt{\delta_{\alpha}^{2}+C_{\alpha}^{2}}}\\
\delta_{\alpha} & = & -\frac{5C_{6}^{(1)}}{81}+\frac{C_{6}^{(2)}}{81}-\frac{2C_{6}^{(3)}}{15}+\frac{16C_{6}^{(4)}}{50625}\\&+&\frac{19C_{6}^{(5)}}{1875}-\frac{9C_{6}^{(6)}}{625}\\
C_{\alpha} & = & \frac{4C_{6}^{(1)}}{81}\sqrt{6}+\frac{16C_{6}^{(2)}}{2025}\sqrt{6}-\frac{8C_{6}^{(3)}}{75}\sqrt{6}\\
 &  & +\frac{16C_{6}^{(4)}}{50625}\sqrt{6}-\frac{16C_{6}^{(5)}}{1875}\sqrt{6}+\frac{36C_{6}^{(6)}}{625}\sqrt{6}
\end{eqnarray*}
and eigenvalues 
\begin{eqnarray*}
\lambda_{3} & = & \frac{11C_{6}^{(1)}}{81}+\frac{53C_{6}^{(2)}}{2025}+\frac{26C_{6}^{(3)}}{75}+\frac{116C_{6}^{(4)}}{50625}+\frac{79C_{6}^{(5)}}{1875}\\&+&\frac{171C_{6}^{(6)}}{625}
  -\sqrt{C_\alpha^{2}+\delta_\alpha^{2}}\\
\lambda_{4} & = & \frac{11C_{6}^{(1)}}{81}+\frac{53C_{6}^{(2)}}{2025}+\frac{26C_{6}^{(3)}}{75}+\frac{116C_{6}^{(4)}}{50625}+\frac{79C_{6}^{(5)}}{1875}\\&+&\frac{171C_{6}^{(6)}}{625} +\sqrt{C_\alpha^{2}+\delta_\alpha^{2}}\\
\lambda_{5} & = & \frac{14C_{6}^{(2)}}{225}+\frac{4C_{6}^{(3)}}{25}+\frac{4C_{6}^{(4)}}{5625}+\frac{38C_{6}^{(5)}}{625}+\frac{156C_{6}^{(6)}}{625}.
\end{eqnarray*}
\subsection{$M=0$}
\begin{figure}[h!]
\begin{centering}
\includegraphics[width=0.35\textwidth]{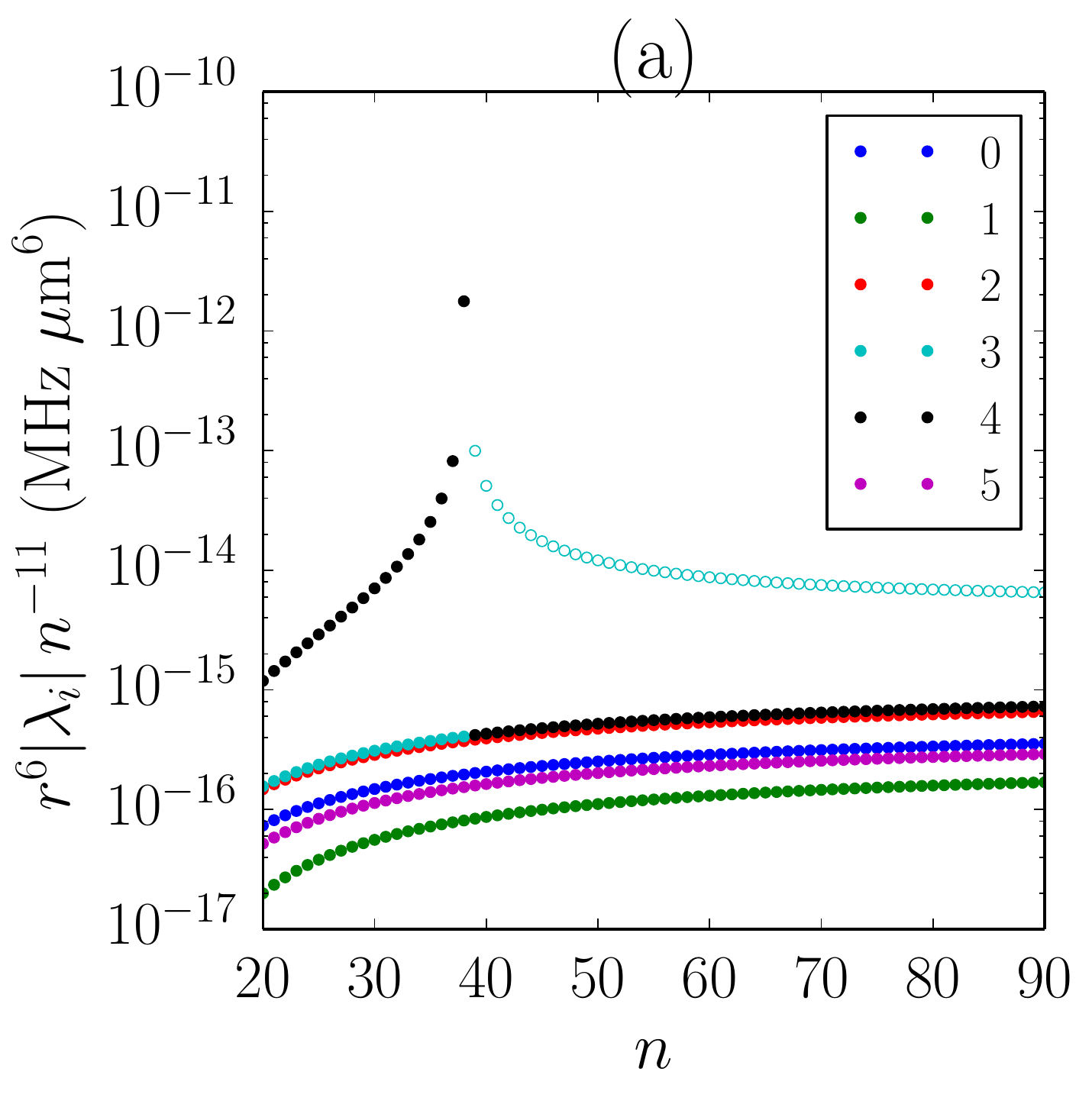}\\
\includegraphics[width=0.35\textwidth]{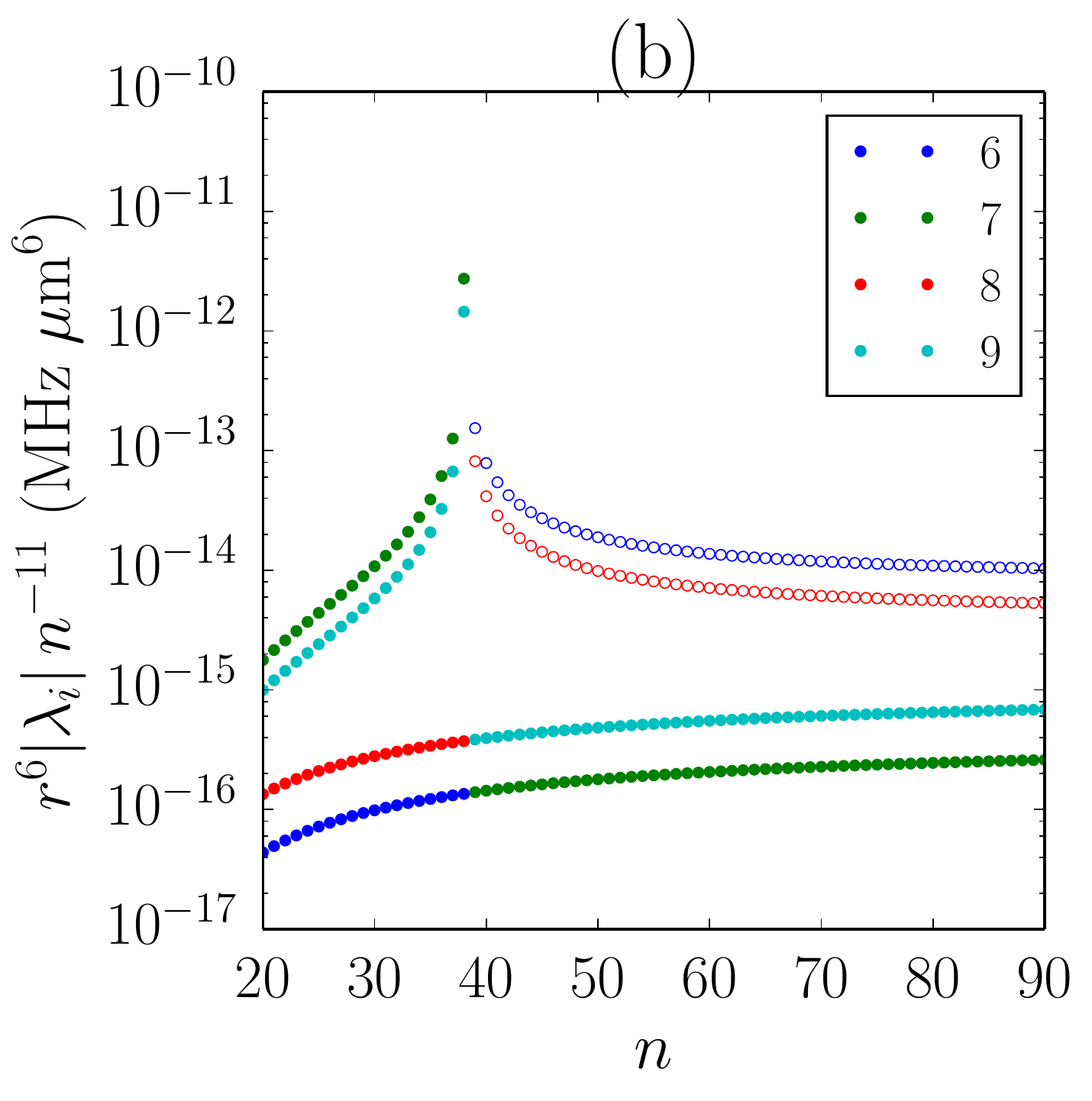} 
\par\end{centering}

\caption{(a) Degenerate and (b) non-degenerate eigenvalues $r^{6}\lambda_{i}$
for Rubidium $P_{3/2}$ states as a function of the principal quantum
number $n$. Empty (full) symbols represent negative (positive) values.\label{fig:lambdavsn_p32} }
\end{figure}

\begin{figure}[h!]
\begin{centering}
\includegraphics[width=0.35\textwidth]{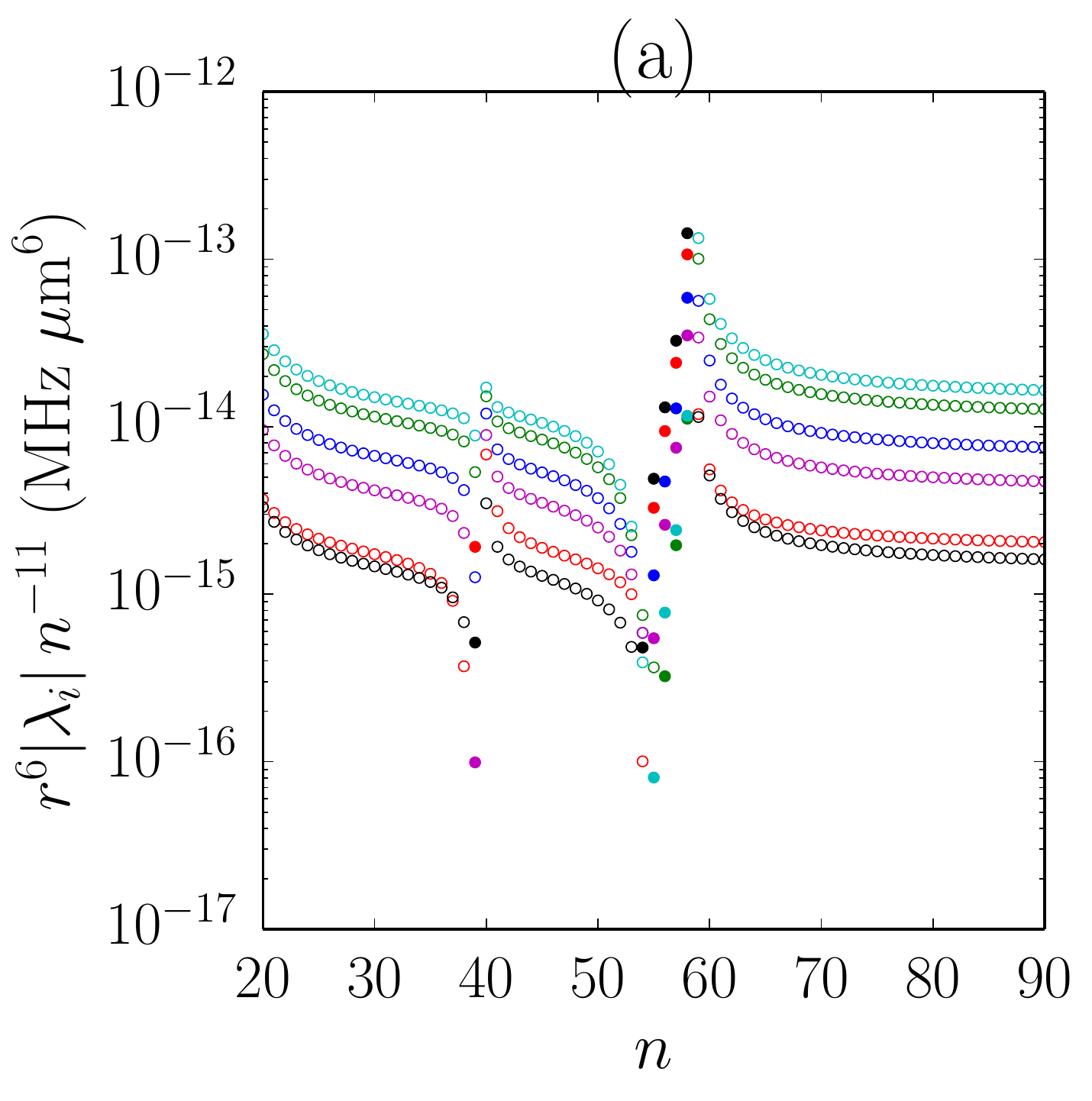}\\
\includegraphics[width=0.35\textwidth]{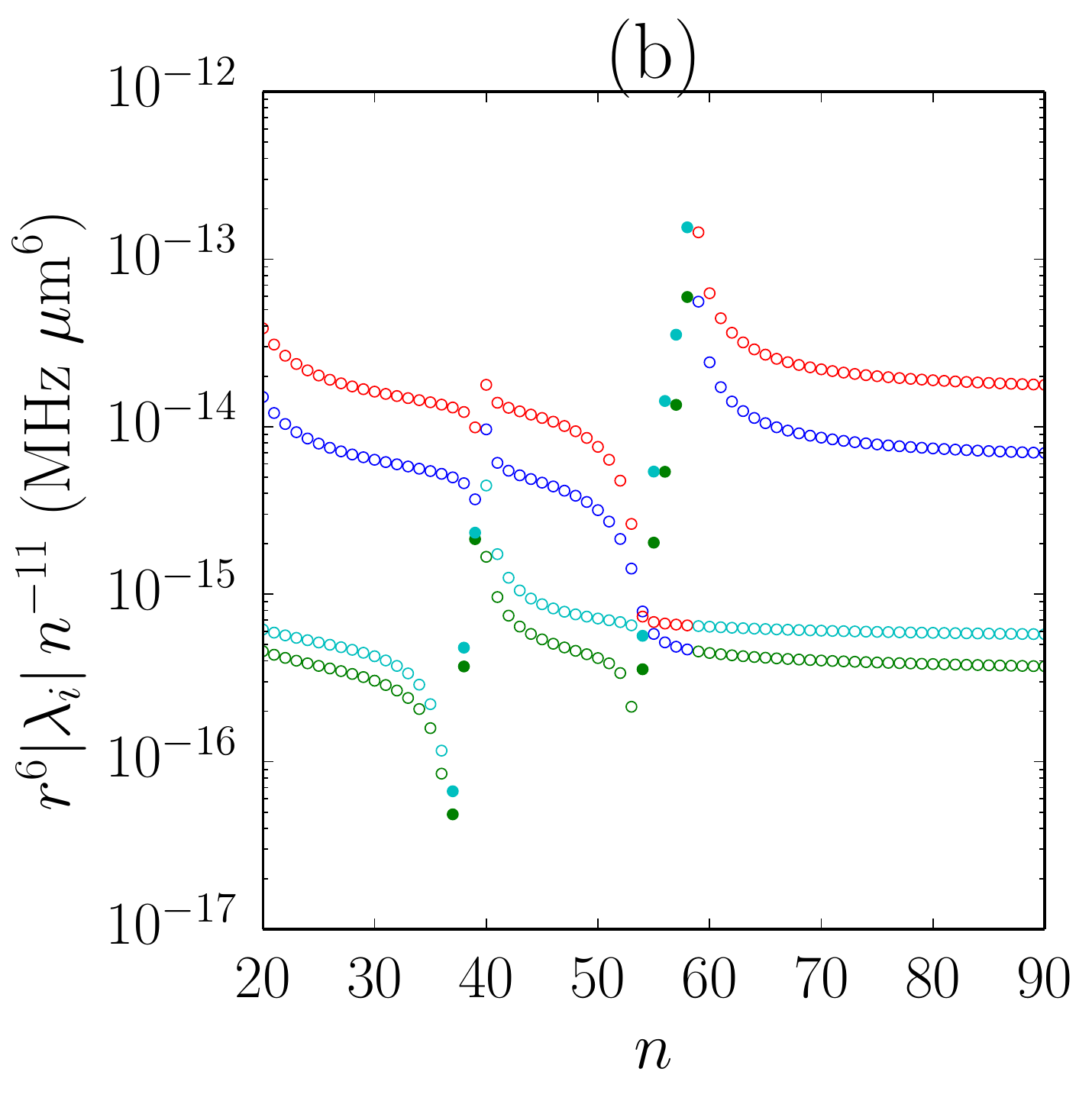} 
\par\end{centering}

\caption{(a) Degenerate and (b) non-degenerate eigenvalues $r^{6}\lambda_{i}$
for Rubidium $D_{3/2}$ states as a function of the principal quantum
number $n$. Empty (full) symbols represent negative (positive) values and the color code is the same as in Fig.~\ref{fig:lambdavsn_p32}.\label{fig:lambdavsn_d32} }
\end{figure}

The last $4$ unknown eigenstates belong to the zero angular momentum subspace: 
\begin{eqnarray*}
|\lambda_{6}\rangle & = & \frac{\cos\beta}{\sqrt{2}}\left[|\mthree,\pthree\rangle+|\pthree,\mthree\rangle\right]+\frac{\sin\beta}{\sqrt{2}}\left[|\mone,\pone\rangle+|\pone,\mone\rangle\right]\\
|\lambda_{7}\rangle & = & -\frac{\sin\beta}{\sqrt{2}}\left[|\mthree,\pthree\rangle+|\pthree,\mthree\rangle\right]+\frac{\cos\beta}{\sqrt{2}}\left[|\mone,\pone\rangle+|\pone,\mone\rangle\right]\\
|\lambda_{8}\rangle & = & \frac{\cos\gamma}{\sqrt{2}}\left[|\mthree,\pthree\rangle-|\pthree,\mthree\rangle\right]+\frac{\sin\gamma}{\sqrt{2}}\left[|\mone,\pone\rangle-|\pone,\mone\rangle\right]\\
|\lambda_{9}\rangle & = & \frac{-\sin\gamma}{\sqrt{2}}\left[|\mthree,\pthree\rangle-|\pthree,\mthree\rangle\right]+\frac{\cos\gamma}{\sqrt{2}}\left[|\mone,\pone\rangle-|\pone,\mone\rangle\right]
\end{eqnarray*}
with 
\begin{eqnarray*}
\cos(2\beta) & = & \frac{\delta_{\beta}}{\sqrt{\delta_{\beta}^{2}+C_{\beta}^{2}}}\\
\sin(2\beta) & = & \frac{C_{\beta}}{\sqrt{\delta_{\beta}^{2}+C_{\beta}^{2}}}\\
\delta_{\beta} & = & -\frac{8C_{6}^{(1)}}{81}-\frac{14C_{6}^{(2)}}{2025}-\frac{8C_{6}^{(3)}}{75}\\&+&\frac{112C_{6}^{(4)}}{50625}+\frac{38C_{6}^{(5)}}{1875}-\frac{48C_{6}^{(6)}}{625}\\
C_{\beta} & = & \frac{5C_{6}^{(1)}}{27}-\frac{16C_{6}^{(2)}}{675}-\frac{2C_{6}^{(3)}}{25}+\frac{56C_{6}^{(4)}}{16875}\\&-&\frac{16C_{6}^{(5)}}{625}+\frac{63C_{6}^{(6)}}{625},
\end{eqnarray*}
\begin{eqnarray*}
\cos(2\gamma) & = & \frac{\delta_{\gamma}}{\sqrt{\delta_{\gamma}^{2}+C_{\gamma}^{2}}}\\
\sin(2\gamma) & = & \frac{C_{\gamma}}{\sqrt{\delta_{\gamma}^{2}+C_{\gamma}^{2}}}\\
\delta_{\gamma} & = & \frac{2C_{6}^{(2)}}{225}-\frac{8C_{6}^{(3)}}{25}+\frac{16C_{6}^{(4)}}{5625}+\frac{2C_{6}^{(5)}}{625}+\frac{24C_{6}^{(6)}}{625}\\
C_{\gamma} & = & \frac{C_{6}^{(1)}}{9}-\frac{2C_{6}^{(3)}}{15}+\frac{8C_{6}^{(4)}}{5625}-\frac{32C_{6}^{(5)}}{1875}+\frac{57C_{6}^{(6)}}{625}
\end{eqnarray*}
and with energies 
\begin{eqnarray*}
\lambda_{6} & = & \frac{17C_{6}^{(1)}}{81}+\frac{2C_{6}^{(2)}}{81}+\frac{2C_{6}^{(3)}}{15}+\frac{248C_{6}^{(4)}}{50625}+\frac{62C_{6}^{(5)}}{1875}+\frac{213C_{6}^{(6)}}{625}\\
 &  & -\sqrt{C_{\beta}^{2}+\delta_{\beta}^{2}}\\
\lambda_{7} & = & \frac{17C_{6}^{(1)}}{81}+\frac{2C_{6}^{(2)}}{81}+\frac{2C_{6}^{(3)}}{15}+\frac{248C_{6}^{(4)}}{50625}+\frac{62C_{6}^{(5)}}{1875}+\frac{213C_{6}^{(6)}}{625}\\
 &  & +\sqrt{C_{\beta}^{2}+\delta_{\beta}^{2}}\\
\lambda_{8} & = & \frac{C_{6}^{(1)}}{9}+\frac{2C_{6}^{(2)}}{225}+\frac{26C_{6}^{(3)}}{75}+\frac{8C_{6}^{(4)}}{1875}+\frac{94C_{6}^{(5)}}{1875}+\frac{141C_{6}^{(6)}}{625}\\
 &  & -\sqrt{C_{\gamma}^{2}+\delta_{\gamma}^{2}}\\
\lambda_{9} & = & \frac{C_{6}^{(1)}}{9}+\frac{2C_{6}^{(2)}}{225}+\frac{26C_{6}^{(3)}}{75}+\frac{8C_{6}^{(4)}}{1875}+\frac{94C_{6}^{(5)}}{1875}+\frac{141C_{6}^{(6)}}{625}\\
 &  & +\sqrt{C_{\gamma}^{2}+\delta_{\gamma}^{2}}.
\end{eqnarray*}
We represent in Fig.~\ref{fig:lambdavsn_p32} the values of the eigenvalues for $P_{3/2}$
states as a function of the principal quantum number $n$. We checked
that the values obtained with these analytical expressions coincide
with the ones obtained by ``brute'' numerical diagonalization of
the vdW Hamiltonian. For $P_{3/2}$ and $n<38$, all the energies
are positive while on the other side of the F\"orster resonance, for
$n>38$, some of the eigenstates become attractive. For $D_{3/2}$
states, the two F\"orster resonances are also visible {[}Fig.~\ref{fig:lambdavsn_d32}{]}.

\section{Effects of the Quadrupole-Quadrupole interactions \label{app:quad}}
In order to show that the vdW interactions are dominant for the range of parameters considered in this work, we now calculate the second term of the multipole expansion corresponding to quadrupole-quadrupole interactions. In a fine-structure manifold ($n,\ell,j$), the matrix elements of the quadrupole-quadrupole Hamiltonian can be written in the form~\cite{Vaillant2012}
\begin{equation}
H_{5} =\frac{C_5}{r^5}  \sum_{m_1,m_2,m_3,m_4}  |m_1, m_2 \rangle D_{22} \langle m_3 ,m_4|  
\end{equation}
with $C_5=\langle n,\ell,j|r^2|n,\ell,j \rangle^2$ is the radial part and $D_{22}$ the angular part.

\begin{figure}[b]
\begin{centering}
\includegraphics[width=0.35\textwidth]{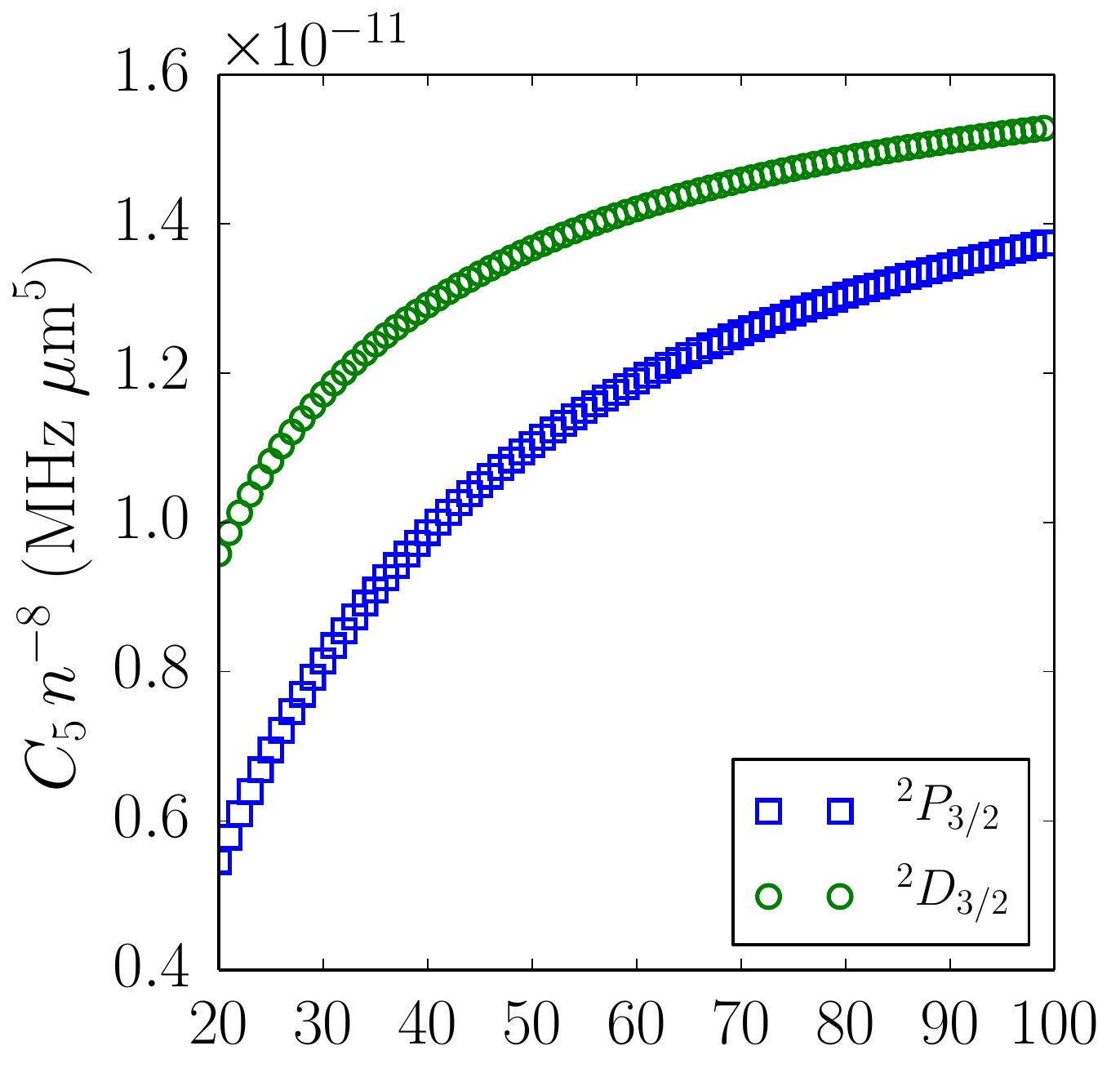}
\par\end{centering}
\caption{$C_5$ coefficient as a function of the principal quantum number for Rubidium $P_{3/2}$ and $D_{3/2}$ atoms. \label{fig:C5} }
\end{figure}

The selection rules of the quadrupole-quadrupole operator impose $j\ge1$ \cite{Vaillant2012}, i.e there are no quadrupole-quadrupole interactions with $S_{1/2}$ and $P_{1/2}$ states. We now consider the case of $P_{3/2}$ and $D_{3/2}$ states. The eigenstates of $H_5$ can be classified in the case $\theta=0$ by values of $M=m_1+m_2$. There are given for $M\ge0$ in Tab.~\ref{tab:H5} while the $C_5$ coefficient, is shown in Fig.~\ref{fig:C5}. Our values, calculated from the model potential~\cite{Marinescu1994}, including spin-orbit effects are in good agreement with previous calculations \cite{Singer2005}. For the parameters of \cite{Barredo2014} the $C_5$ coefficient is $30.5$ GHz $\mu\mathrm{m}^5$. Given that $C_6 = -8860$ GHz $\mu\mathrm{m}^6$ and for distances of the order of a few micrometers, the quadrupole-quadrupole interactions are therefore negligible.
\begin{table}[h]
\begin{ruledtabular}
\begin{tabular}{ccc}
$M$ & $\lambda r^5/C_5$ & $|\lambda\rangle$  \tabularnewline
3    & $\tfrac{6}{25}$          & $|\pthree,\pthree\rangle$  \tabularnewline
2    & $-\tfrac{14}{25}$          & $\tfrac{1}{\sqrt{2}}\left(|\pthree,\pone\rangle+|\pthree,\pone\rangle\right)$  \tabularnewline
2    & $\tfrac{2}{25}$          & $\tfrac{1}{\sqrt{2}}\left(|\pthree,\pone\rangle-|\pthree,\pone\rangle\right)$  \tabularnewline
1    & $\tfrac{6}{25}$          & $|\pone,\pone\rangle$  \tabularnewline
1    & $-\tfrac{4}{25}$          & $\tfrac{1}{\sqrt{2}}\left(|\pthree,\mone\rangle+|\mone,\pthree\rangle\right)$  \tabularnewline
1    & $-\tfrac{8}{25}$          & $\tfrac{1}{\sqrt{2}}\left(|\pthree,\mone\rangle-|\mone,\pthree\rangle\right)$  \tabularnewline
0    & $\tfrac{16}{25}$          & $\tfrac{1}{2}\left(|\pthree,\mthree\rangle+|\mthree,\pthree\rangle+|\pone,\mone\rangle+|\mone,\pone\rangle \right)$  \tabularnewline
0    & -$\tfrac{4}{25}$          & $\tfrac{1}{2}\left(|\pthree,\mthree\rangle+|\mthree,\pthree\rangle-|\pone,\mone\rangle-|\mone,\pone\rangle \right)$  \tabularnewline
0    & $0$          & $\tfrac{1}{2}\left(|\pthree,\mthree\rangle-|\mthree,\pthree\rangle-|\pone,\mone\rangle+|\mone,\pone\rangle \right)$  \tabularnewline
0    & $\tfrac{12}{25}$          & $\tfrac{1}{2}\left(|\pthree,\mthree\rangle-|\mthree,\pthree\rangle+|\pone,\mone\rangle-|\mone,\pone\rangle \right)$  \tabularnewline
\end{tabular}\end{ruledtabular}
\caption{Eigenstates and eigenvalues of the quadrupole-quadrupole Hamiltonian for $j=3/2$.\label{tab:H5}}
\end{table}

\clearpage
\bibliographystyle{apsrev4-1}
\bibliography{mixing}

\begin{thebibliography}{35}%
\makeatletter
\providecommand \@ifxundefined [1]{%
 \@ifx{#1\undefined}
}%
\providecommand \@ifnum [1]{%
 \ifnum #1\expandafter \@firstoftwo
 \else \expandafter \@secondoftwo
 \fi
}%
\providecommand \@ifx [1]{%
 \ifx #1\expandafter \@firstoftwo
 \else \expandafter \@secondoftwo
 \fi
}%
\providecommand \natexlab [1]{#1}%
\providecommand \enquote  [1]{``#1''}%
\providecommand \bibnamefont  [1]{#1}%
\providecommand \bibfnamefont [1]{#1}%
\providecommand \citenamefont [1]{#1}%
\providecommand \href@noop [0]{\@secondoftwo}%
\providecommand \href [0]{\begingroup \@sanitize@url \@href}%
\providecommand \@href[1]{\@@startlink{#1}\@@href}%
\providecommand \@@href[1]{\endgroup#1\@@endlink}%
\providecommand \@sanitize@url [0]{\catcode `\\12\catcode `\$12\catcode
  `\&12\catcode `\#12\catcode `\^12\catcode `\_12\catcode `\%12\relax}%
\providecommand \@@startlink[1]{}%
\providecommand \@@endlink[0]{}%
\providecommand \url  [0]{\begingroup\@sanitize@url \@url }%
\providecommand \@url [1]{\endgroup\@href {#1}{\urlprefix }}%
\providecommand \urlprefix  [0]{URL }%
\providecommand \Eprint [0]{\href }%
\providecommand \doibase [0]{http://dx.doi.org/}%
\providecommand \selectlanguage [0]{\@gobble}%
\providecommand \bibinfo  [0]{\@secondoftwo}%
\providecommand \bibfield  [0]{\@secondoftwo}%
\providecommand \translation [1]{[#1]}%
\providecommand \BibitemOpen [0]{}%
\providecommand \bibitemStop [0]{}%
\providecommand \bibitemNoStop [0]{.\EOS\space}%
\providecommand \EOS [0]{\spacefactor3000\relax}%
\providecommand \BibitemShut  [1]{\csname bibitem#1\endcsname}%
\let\auto@bib@innerbib\@empty
\bibitem [{\citenamefont {Gallagher}(2005)}]{gallagher2005rydberg}%
  \BibitemOpen
  \bibfield  {author} {\bibinfo {author} {\bibfnamefont {T.~F.}\ \bibnamefont
  {Gallagher}},\ }\href {http://books.google.at/books?id=8JIpEhHWT-cC} {\emph
  {\bibinfo {title} {{Rydberg Atoms}}}},\ Cambridge Monographs on Atomic,
  Molecular and Chemical Physics\ (\bibinfo  {publisher} {Cambridge University
  Press},\ \bibinfo {year} {2005})\BibitemShut {NoStop}%
\bibitem [{\citenamefont {L\"{o}w}\ \emph {et~al.}(2012)\citenamefont
  {L\"{o}w}, \citenamefont {Weimer}, \citenamefont {Nipper}, \citenamefont
  {Balewski}, \citenamefont {Butscher}, \citenamefont {B\"{u}chler},\ and\
  \citenamefont {Pfau}}]{Low2012}%
  \BibitemOpen
  \bibfield  {author} {\bibinfo {author} {\bibfnamefont {R.}~\bibnamefont
  {L\"{o}w}}, \bibinfo {author} {\bibfnamefont {H.}~\bibnamefont {Weimer}},
  \bibinfo {author} {\bibfnamefont {J.}~\bibnamefont {Nipper}}, \bibinfo
  {author} {\bibfnamefont {J.~B.}\ \bibnamefont {Balewski}}, \bibinfo {author}
  {\bibfnamefont {B.}~\bibnamefont {Butscher}}, \bibinfo {author}
  {\bibfnamefont {H.~P.}\ \bibnamefont {B\"{u}chler}}, \ and\ \bibinfo {author}
  {\bibfnamefont {T.}~\bibnamefont {Pfau}},\ }\href
  {http://stacks.iop.org/0953-4075/45/i=11/a=113001?key=crossref.12ca63182fe7b970c8c27db2e22c4fb2}
  {\bibfield  {journal} {\bibinfo  {journal} {Journal of Physics B: Atomic,
  Molecular and Optical Physics}\ }\textbf {\bibinfo {volume} {45}},\ \bibinfo
  {pages} {113001} (\bibinfo {year} {2012})}\BibitemShut {NoStop}%
\bibitem [{\citenamefont {Saffman}\ \emph {et~al.}(2010)\citenamefont
  {Saffman}, \citenamefont {Walker},\ and\ \citenamefont
  {M\o~lmer}}]{Saffman2010}%
  \BibitemOpen
  \bibfield  {author} {\bibinfo {author} {\bibfnamefont {M.}~\bibnamefont
  {Saffman}}, \bibinfo {author} {\bibfnamefont {T.~G.}\ \bibnamefont {Walker}},
  \ and\ \bibinfo {author} {\bibfnamefont {K.}~\bibnamefont {M\o~lmer}},\
  }\href {\doibase 10.1103/RevModPhys.82.2313} {\bibfield  {journal} {\bibinfo
  {journal} {Reviews of Modern Physics}\ }\textbf {\bibinfo {volume} {82}},\
  \bibinfo {pages} {2313} (\bibinfo {year} {2010})}\BibitemShut {NoStop}%
\bibitem [{\citenamefont {Jaksch}\ \emph {et~al.}(2000)\citenamefont {Jaksch},
  \citenamefont {Cirac}, \citenamefont {Zoller}, \citenamefont {Rolston},
  \citenamefont {C\^{o}t\'{e}},\ and\ \citenamefont {Lukin}}]{Jaksch2000}%
  \BibitemOpen
  \bibfield  {author} {\bibinfo {author} {\bibfnamefont {D.}~\bibnamefont
  {Jaksch}}, \bibinfo {author} {\bibfnamefont {J.~I.}\ \bibnamefont {Cirac}},
  \bibinfo {author} {\bibfnamefont {P.}~\bibnamefont {Zoller}}, \bibinfo
  {author} {\bibfnamefont {S.~L.}\ \bibnamefont {Rolston}}, \bibinfo {author}
  {\bibfnamefont {R.}~\bibnamefont {C\^{o}t\'{e}}}, \ and\ \bibinfo {author}
  {\bibfnamefont {M.~D.}\ \bibnamefont {Lukin}},\ }\href {\doibase
  10.1103/PhysRevLett.85.2208} {\bibfield  {journal} {\bibinfo  {journal}
  {Phys. Rev. Lett.}\ }\textbf {\bibinfo {volume} {85}},\ \bibinfo {pages}
  {2208} (\bibinfo {year} {2000})}\BibitemShut {NoStop}%
\bibitem [{\citenamefont {Lukin}\ \emph {et~al.}(2001)\citenamefont {Lukin},
  \citenamefont {Fleischhauer}, \citenamefont {Cote}, \citenamefont {Duan},
  \citenamefont {Jaksch}, \citenamefont {Cirac},\ and\ \citenamefont
  {Zoller}}]{Lukin2001}%
  \BibitemOpen
  \bibfield  {author} {\bibinfo {author} {\bibfnamefont {M.~D.}\ \bibnamefont
  {Lukin}}, \bibinfo {author} {\bibfnamefont {M.}~\bibnamefont {Fleischhauer}},
  \bibinfo {author} {\bibfnamefont {R.}~\bibnamefont {Cote}}, \bibinfo {author}
  {\bibfnamefont {L.~M.}\ \bibnamefont {Duan}}, \bibinfo {author}
  {\bibfnamefont {D.}~\bibnamefont {Jaksch}}, \bibinfo {author} {\bibfnamefont
  {J.~I.}\ \bibnamefont {Cirac}}, \ and\ \bibinfo {author} {\bibfnamefont
  {P.}~\bibnamefont {Zoller}},\ }\href {\doibase 10.1103/PhysRevLett.87.037901}
  {\bibfield  {journal} {\bibinfo  {journal} {Phys. Rev. Lett.}\ }\textbf
  {\bibinfo {volume} {87}},\ \bibinfo {pages} {37901} (\bibinfo {year}
  {2001})}\BibitemShut {NoStop}%
\bibitem [{\citenamefont {Tong}\ \emph {et~al.}(2004)\citenamefont {Tong},
  \citenamefont {Farooqi}, \citenamefont {Stanojevic}, \citenamefont
  {Krishnan}, \citenamefont {Zhang}, \citenamefont {C\^{o}t\'{e}},
  \citenamefont {Eyler},\ and\ \citenamefont {Gould}}]{Tong2004}%
  \BibitemOpen
  \bibfield  {author} {\bibinfo {author} {\bibfnamefont {D.}~\bibnamefont
  {Tong}}, \bibinfo {author} {\bibfnamefont {S.~M.}\ \bibnamefont {Farooqi}},
  \bibinfo {author} {\bibfnamefont {J.}~\bibnamefont {Stanojevic}}, \bibinfo
  {author} {\bibfnamefont {S.}~\bibnamefont {Krishnan}}, \bibinfo {author}
  {\bibfnamefont {Y.~P.}\ \bibnamefont {Zhang}}, \bibinfo {author}
  {\bibfnamefont {R.}~\bibnamefont {C\^{o}t\'{e}}}, \bibinfo {author}
  {\bibfnamefont {E.~E.}\ \bibnamefont {Eyler}}, \ and\ \bibinfo {author}
  {\bibfnamefont {P.~L.}\ \bibnamefont {Gould}},\ }\href {\doibase
  10.1103/PhysRevLett.93.063001} {\bibfield  {journal} {\bibinfo  {journal}
  {Physical Review Letters}\ }\textbf {\bibinfo {volume} {93}},\ \bibinfo
  {pages} {063001} (\bibinfo {year} {2004})}\BibitemShut {NoStop}%
\bibitem [{\citenamefont {Singer}\ \emph {et~al.}(2004)\citenamefont {Singer},
  \citenamefont {Reetz-Lamour}, \citenamefont {Amthor}, \citenamefont
  {Marcassa},\ and\ \citenamefont {Weidem\"{u}ller}}]{Singer2004}%
  \BibitemOpen
  \bibfield  {author} {\bibinfo {author} {\bibfnamefont {K.}~\bibnamefont
  {Singer}}, \bibinfo {author} {\bibfnamefont {M.}~\bibnamefont
  {Reetz-Lamour}}, \bibinfo {author} {\bibfnamefont {T.}~\bibnamefont
  {Amthor}}, \bibinfo {author} {\bibfnamefont {L.~G.}\ \bibnamefont
  {Marcassa}}, \ and\ \bibinfo {author} {\bibfnamefont {M.}~\bibnamefont
  {Weidem\"{u}ller}},\ }\href {\doibase 10.1103/PhysRevLett.93.163001}
  {\bibfield  {journal} {\bibinfo  {journal} {Physical Review Letters}\
  }\textbf {\bibinfo {volume} {93}},\ \bibinfo {pages} {163001} (\bibinfo
  {year} {2004})}\BibitemShut {NoStop}%
\bibitem [{\citenamefont {Ga\"{e}tan}\ \emph {et~al.}(2009)\citenamefont
  {Ga\"{e}tan}, \citenamefont {Miroshnychenko}, \citenamefont {Wilk},
  \citenamefont {Chotia}, \citenamefont {Viteau}, \citenamefont {Comparat},
  \citenamefont {Pillet}, \citenamefont {Browaeys},\ and\ \citenamefont
  {Grangier}}]{Gaetan2009}%
  \BibitemOpen
  \bibfield  {author} {\bibinfo {author} {\bibfnamefont {A.}~\bibnamefont
  {Ga\"{e}tan}}, \bibinfo {author} {\bibfnamefont {Y.}~\bibnamefont
  {Miroshnychenko}}, \bibinfo {author} {\bibfnamefont {T.}~\bibnamefont
  {Wilk}}, \bibinfo {author} {\bibfnamefont {A.}~\bibnamefont {Chotia}},
  \bibinfo {author} {\bibfnamefont {M.}~\bibnamefont {Viteau}}, \bibinfo
  {author} {\bibfnamefont {D.}~\bibnamefont {Comparat}}, \bibinfo {author}
  {\bibfnamefont {P.}~\bibnamefont {Pillet}}, \bibinfo {author} {\bibfnamefont
  {A.}~\bibnamefont {Browaeys}}, \ and\ \bibinfo {author} {\bibfnamefont
  {P.}~\bibnamefont {Grangier}},\ }\href {\doibase 10.1038/nphys1183}
  {\bibfield  {journal} {\bibinfo  {journal} {Nature Physics}\ }\textbf
  {\bibinfo {volume} {5}},\ \bibinfo {pages} {115} (\bibinfo {year}
  {2009})}\BibitemShut {NoStop}%
\bibitem [{\citenamefont {Urban}\ \emph {et~al.}(2009)\citenamefont {Urban},
  \citenamefont {Johnson}, \citenamefont {Henage}, \citenamefont {Isenhower},
  \citenamefont {Yavuz}, \citenamefont {Walker},\ and\ \citenamefont
  {Saffman}}]{Urban2009}%
  \BibitemOpen
  \bibfield  {author} {\bibinfo {author} {\bibfnamefont {E.}~\bibnamefont
  {Urban}}, \bibinfo {author} {\bibfnamefont {T.~a.}\ \bibnamefont {Johnson}},
  \bibinfo {author} {\bibfnamefont {T.}~\bibnamefont {Henage}}, \bibinfo
  {author} {\bibfnamefont {L.}~\bibnamefont {Isenhower}}, \bibinfo {author}
  {\bibfnamefont {D.~D.}\ \bibnamefont {Yavuz}}, \bibinfo {author}
  {\bibfnamefont {T.~G.}\ \bibnamefont {Walker}}, \ and\ \bibinfo {author}
  {\bibfnamefont {M.}~\bibnamefont {Saffman}},\ }\href {\doibase
  10.1038/nphys1178} {\bibfield  {journal} {\bibinfo  {journal} {Nature
  Physics}\ }\textbf {\bibinfo {volume} {5}},\ \bibinfo {pages} {110} (\bibinfo
  {year} {2009})}\BibitemShut {NoStop}%
\bibitem [{\citenamefont {Ebert}\ \emph {et~al.}(2014)\citenamefont {Ebert},
  \citenamefont {Gill}, \citenamefont {Gibbons}, \citenamefont {Zhang},
  \citenamefont {Saffman},\ and\ \citenamefont {Walker}}]{Ebert2014}%
  \BibitemOpen
  \bibfield  {author} {\bibinfo {author} {\bibfnamefont {M.}~\bibnamefont
  {Ebert}}, \bibinfo {author} {\bibfnamefont {A.}~\bibnamefont {Gill}},
  \bibinfo {author} {\bibfnamefont {M.}~\bibnamefont {Gibbons}}, \bibinfo
  {author} {\bibfnamefont {X.}~\bibnamefont {Zhang}}, \bibinfo {author}
  {\bibfnamefont {M.}~\bibnamefont {Saffman}}, \ and\ \bibinfo {author}
  {\bibfnamefont {T.~G.}\ \bibnamefont {Walker}},\ }\href {\doibase
  10.1103/PhysRevLett.112.043602} {\bibfield  {journal} {\bibinfo  {journal}
  {Physical Review Letters}\ }\textbf {\bibinfo {volume} {112}},\ \bibinfo
  {pages} {043602} (\bibinfo {year} {2014})}\BibitemShut {NoStop}%
\bibitem [{\citenamefont {Barredo}\ \emph {et~al.}(2014)\citenamefont
  {Barredo}, \citenamefont {Ravets}, \citenamefont {Labuhn}, \citenamefont
  {B\'{e}guin}, \citenamefont {Vernier}, \citenamefont {Nogrette},
  \citenamefont {Lahaye},\ and\ \citenamefont {Browaeys}}]{Barredo2014}%
  \BibitemOpen
  \bibfield  {author} {\bibinfo {author} {\bibfnamefont {D.}~\bibnamefont
  {Barredo}}, \bibinfo {author} {\bibfnamefont {S.}~\bibnamefont {Ravets}},
  \bibinfo {author} {\bibfnamefont {H.}~\bibnamefont {Labuhn}}, \bibinfo
  {author} {\bibfnamefont {L.}~\bibnamefont {B\'{e}guin}}, \bibinfo {author}
  {\bibfnamefont {A.}~\bibnamefont {Vernier}}, \bibinfo {author} {\bibfnamefont
  {F.}~\bibnamefont {Nogrette}}, \bibinfo {author} {\bibfnamefont
  {T.}~\bibnamefont {Lahaye}}, \ and\ \bibinfo {author} {\bibfnamefont
  {A.}~\bibnamefont {Browaeys}},\ }\href {\doibase
  10.1103/PhysRevLett.112.183002} {\bibfield  {journal} {\bibinfo  {journal}
  {Phys. Rev. Lett.}\ }\textbf {\bibinfo {volume} {112}},\ \bibinfo {pages}
  {183002} (\bibinfo {year} {2014})}\BibitemShut {NoStop}%
\bibitem [{\citenamefont {Schau\ss}\ \emph {et~al.}()\citenamefont {Schau\ss},
  \citenamefont {Zeiher}, \citenamefont {Fukuhara}, \citenamefont {Hild},
  \citenamefont {Cheneau}, \citenamefont {Macr\`{\i}}, \citenamefont {Pohl},
  \citenamefont {Bloch},\ and\ \citenamefont {Gross}}]{Schauss2014}%
  \BibitemOpen
  \bibfield  {author} {\bibinfo {author} {\bibfnamefont {P.}~\bibnamefont
  {Schau\ss}}, \bibinfo {author} {\bibfnamefont {J.}~\bibnamefont {Zeiher}},
  \bibinfo {author} {\bibfnamefont {T.}~\bibnamefont {Fukuhara}}, \bibinfo
  {author} {\bibfnamefont {S.}~\bibnamefont {Hild}}, \bibinfo {author}
  {\bibfnamefont {M.}~\bibnamefont {Cheneau}}, \bibinfo {author} {\bibfnamefont
  {T.}~\bibnamefont {Macr\`{\i}}}, \bibinfo {author} {\bibfnamefont
  {T.}~\bibnamefont {Pohl}}, \bibinfo {author} {\bibfnamefont {I.}~\bibnamefont
  {Bloch}}, \ and\ \bibinfo {author} {\bibfnamefont {C.}~\bibnamefont
  {Gross}},\ }\href {http://arxiv.org/abs/1404.0980} {\ }\Eprint
  {http://arxiv.org/abs/1404.0980} {arXiv:1404.0980} \BibitemShut {NoStop}%
\bibitem [{\citenamefont {Hankin}\ \emph {et~al.}(2014)\citenamefont {Hankin},
  \citenamefont {Jau}, \citenamefont {Parazzoli}, \citenamefont {Chou},
  \citenamefont {Armstrong}, \citenamefont {Landahl},\ and\ \citenamefont
  {Biedermann}}]{Hankin2014}%
  \BibitemOpen
  \bibfield  {author} {\bibinfo {author} {\bibfnamefont {a.~M.}\ \bibnamefont
  {Hankin}}, \bibinfo {author} {\bibfnamefont {Y.-Y.}\ \bibnamefont {Jau}},
  \bibinfo {author} {\bibfnamefont {L.~P.}\ \bibnamefont {Parazzoli}}, \bibinfo
  {author} {\bibfnamefont {C.~W.}\ \bibnamefont {Chou}}, \bibinfo {author}
  {\bibfnamefont {D.~J.}\ \bibnamefont {Armstrong}}, \bibinfo {author}
  {\bibfnamefont {a.~J.}\ \bibnamefont {Landahl}}, \ and\ \bibinfo {author}
  {\bibfnamefont {G.~W.}\ \bibnamefont {Biedermann}},\ }\href {\doibase
  10.1103/PhysRevA.89.033416} {\bibfield  {journal} {\bibinfo  {journal}
  {Physical Review A}\ }\textbf {\bibinfo {volume} {89}},\ \bibinfo {pages}
  {033416} (\bibinfo {year} {2014})}\BibitemShut {NoStop}%
\bibitem [{\citenamefont {Viteau}\ \emph {et~al.}(2011)\citenamefont {Viteau},
  \citenamefont {Bason}, \citenamefont {Radogostowicz}, \citenamefont
  {Malossi}, \citenamefont {Ciampini}, \citenamefont {Morsch},\ and\
  \citenamefont {Arimondo}}]{Viteau2011}%
  \BibitemOpen
  \bibfield  {author} {\bibinfo {author} {\bibfnamefont {M.}~\bibnamefont
  {Viteau}}, \bibinfo {author} {\bibfnamefont {M.~G.}\ \bibnamefont {Bason}},
  \bibinfo {author} {\bibfnamefont {J.}~\bibnamefont {Radogostowicz}}, \bibinfo
  {author} {\bibfnamefont {N.}~\bibnamefont {Malossi}}, \bibinfo {author}
  {\bibfnamefont {D.}~\bibnamefont {Ciampini}}, \bibinfo {author}
  {\bibfnamefont {O.}~\bibnamefont {Morsch}}, \ and\ \bibinfo {author}
  {\bibfnamefont {E.}~\bibnamefont {Arimondo}},\ }\href {\doibase
  10.1103/PhysRevLett.107.060402} {\bibfield  {journal} {\bibinfo  {journal}
  {Physical Review Letters}\ }\textbf {\bibinfo {volume} {107}},\ \bibinfo
  {pages} {060402} (\bibinfo {year} {2011})}\BibitemShut {NoStop}%
\bibitem [{\citenamefont {Reinhard}\ \emph {et~al.}(2007)\citenamefont
  {Reinhard}, \citenamefont {Liebisch}, \citenamefont {Knuffman},\ and\
  \citenamefont {Raithel}}]{Reinhard2007}%
  \BibitemOpen
  \bibfield  {author} {\bibinfo {author} {\bibfnamefont {A.}~\bibnamefont
  {Reinhard}}, \bibinfo {author} {\bibfnamefont {T.}~\bibnamefont {Liebisch}},
  \bibinfo {author} {\bibfnamefont {B.}~\bibnamefont {Knuffman}}, \ and\
  \bibinfo {author} {\bibfnamefont {G.}~\bibnamefont {Raithel}},\ }\href
  {\doibase 10.1103/PhysRevA.75.032712} {\bibfield  {journal} {\bibinfo
  {journal} {Physical Review A}\ }\textbf {\bibinfo {volume} {75}},\ \bibinfo
  {pages} {032712} (\bibinfo {year} {2007})}\BibitemShut {NoStop}%
\bibitem [{\citenamefont {Walker}\ and\ \citenamefont
  {Saffman}(2008)}]{Walker2008}%
  \BibitemOpen
  \bibfield  {author} {\bibinfo {author} {\bibfnamefont {T.}~\bibnamefont
  {Walker}}\ and\ \bibinfo {author} {\bibfnamefont {M.}~\bibnamefont
  {Saffman}},\ }\href {\doibase 10.1103/PhysRevA.77.032723} {\bibfield
  {journal} {\bibinfo  {journal} {Physical Review A}\ }\textbf {\bibinfo
  {volume} {77}},\ \bibinfo {pages} {032723} (\bibinfo {year}
  {2008})}\BibitemShut {NoStop}%
\bibitem [{\citenamefont {Glaetzle}\ \emph {et~al.}()\citenamefont {Glaetzle},
  \citenamefont {Dalmonte}, \citenamefont {Nath}, \citenamefont {Gross},
  \citenamefont {Bloch},\ and\ \citenamefont {Zoller}}]{Glaetzle2014}%
  \BibitemOpen
  \bibfield  {author} {\bibinfo {author} {\bibfnamefont {A.~W.}\ \bibnamefont
  {Glaetzle}}, \bibinfo {author} {\bibfnamefont {M.}~\bibnamefont {Dalmonte}},
  \bibinfo {author} {\bibfnamefont {R.}~\bibnamefont {Nath}}, \bibinfo {author}
  {\bibfnamefont {C.}~\bibnamefont {Gross}}, \bibinfo {author} {\bibfnamefont
  {I.}~\bibnamefont {Bloch}}, \ and\ \bibinfo {author} {\bibfnamefont
  {P.}~\bibnamefont {Zoller}},\ }\href {http://arxiv.org/abs/1410.3388v1
  http://arxiv.org/abs/1410.3388} {\ }\Eprint {http://arxiv.org/abs/1410.3388}
  {arXiv:1410.3388} \BibitemShut {NoStop}%
\bibitem [{Note1()}]{Note1}%
  \BibitemOpen
  \bibinfo {note} {These quantities can be interpreted as the Born-Oppenheimer
  energies of the vdW interactions}\BibitemShut {NoStop}%
\bibitem [{Note2()}]{Note2}%
  \BibitemOpen
  \bibinfo {note} {Given their negligible hyperfine splitting \cite
  {Saffman2010,Low2012}, we can describe the Rydberg states using the basis
  associated to the fine-structure, thus considering that the nuclear spin
  behaves as a spectator.}\BibitemShut {Stop}%
\bibitem [{\citenamefont {Glaetzle}\ \emph {et~al.}(2012)\citenamefont
  {Glaetzle}, \citenamefont {Nath}, \citenamefont {Zhao}, \citenamefont
  {Pupillo},\ and\ \citenamefont {Zoller}}]{Glaetzle2012}%
  \BibitemOpen
  \bibfield  {author} {\bibinfo {author} {\bibfnamefont {A.~W.}\ \bibnamefont
  {Glaetzle}}, \bibinfo {author} {\bibfnamefont {R.}~\bibnamefont {Nath}},
  \bibinfo {author} {\bibfnamefont {B.}~\bibnamefont {Zhao}}, \bibinfo {author}
  {\bibfnamefont {G.}~\bibnamefont {Pupillo}}, \ and\ \bibinfo {author}
  {\bibfnamefont {P.}~\bibnamefont {Zoller}},\ }\href {\doibase
  10.1103/PhysRevA.86.043403} {\bibfield  {journal} {\bibinfo  {journal}
  {Physical Review A}\ }\textbf {\bibinfo {volume} {86}},\ \bibinfo {pages}
  {043403} (\bibinfo {year} {2012})}\BibitemShut {NoStop}%
\bibitem [{\citenamefont {Li}\ \emph {et~al.}(2013)\citenamefont {Li},
  \citenamefont {Ates},\ and\ \citenamefont {Lesanovsky}}]{Li2013}%
  \BibitemOpen
  \bibfield  {author} {\bibinfo {author} {\bibfnamefont {W.}~\bibnamefont
  {Li}}, \bibinfo {author} {\bibfnamefont {C.}~\bibnamefont {Ates}}, \ and\
  \bibinfo {author} {\bibfnamefont {I.}~\bibnamefont {Lesanovsky}},\ }\href
  {\doibase 10.1103/PhysRevLett.110.213005} {\bibfield  {journal} {\bibinfo
  {journal} {Physical Review Letters}\ }\textbf {\bibinfo {volume} {110}},\
  \bibinfo {pages} {213005} (\bibinfo {year} {2013})}\BibitemShut {NoStop}%
\bibitem [{\citenamefont {Beterov}\ \emph {et~al.}(2009)\citenamefont
  {Beterov}, \citenamefont {Ryabtsev}, \citenamefont {Tretyakov},\ and\
  \citenamefont {Entin}}]{Beterov2009}%
  \BibitemOpen
  \bibfield  {author} {\bibinfo {author} {\bibfnamefont {I.}~\bibnamefont
  {Beterov}}, \bibinfo {author} {\bibfnamefont {I.}~\bibnamefont {Ryabtsev}},
  \bibinfo {author} {\bibfnamefont {D.}~\bibnamefont {Tretyakov}}, \ and\
  \bibinfo {author} {\bibfnamefont {V.}~\bibnamefont {Entin}},\ }\href
  {\doibase 10.1103/PhysRevA.79.052504} {\bibfield  {journal} {\bibinfo
  {journal} {Physical Review A}\ }\textbf {\bibinfo {volume} {79}},\ \bibinfo
  {pages} {052504} (\bibinfo {year} {2009})}\BibitemShut {NoStop}%
\bibitem [{\citenamefont {Ates}\ \emph {et~al.}(2007)\citenamefont {Ates},
  \citenamefont {Pohl}, \citenamefont {Pattard},\ and\ \citenamefont
  {Rost}}]{Ates2007}%
  \BibitemOpen
  \bibfield  {author} {\bibinfo {author} {\bibfnamefont {C.}~\bibnamefont
  {Ates}}, \bibinfo {author} {\bibfnamefont {T.}~\bibnamefont {Pohl}}, \bibinfo
  {author} {\bibfnamefont {T.}~\bibnamefont {Pattard}}, \ and\ \bibinfo
  {author} {\bibfnamefont {J.}~\bibnamefont {Rost}},\ }\href {\doibase
  10.1103/PhysRevLett.98.023002} {\bibfield  {journal} {\bibinfo  {journal}
  {Physical Review Letters}\ }\textbf {\bibinfo {volume} {98}},\ \bibinfo
  {pages} {023002} (\bibinfo {year} {2007})}\BibitemShut {NoStop}%
\bibitem [{\citenamefont {B\'{e}guin}\ \emph {et~al.}(2013)\citenamefont
  {B\'{e}guin}, \citenamefont {Vernier}, \citenamefont {Chicireanu},
  \citenamefont {Lahaye},\ and\ \citenamefont {Browaeys}}]{Beguin2013}%
  \BibitemOpen
  \bibfield  {author} {\bibinfo {author} {\bibfnamefont {L.}~\bibnamefont
  {B\'{e}guin}}, \bibinfo {author} {\bibfnamefont {A.}~\bibnamefont {Vernier}},
  \bibinfo {author} {\bibfnamefont {R.}~\bibnamefont {Chicireanu}}, \bibinfo
  {author} {\bibfnamefont {T.}~\bibnamefont {Lahaye}}, \ and\ \bibinfo {author}
  {\bibfnamefont {A.}~\bibnamefont {Browaeys}},\ }\href {\doibase
  10.1103/PhysRevLett.110.263201} {\bibfield  {journal} {\bibinfo  {journal}
  {Physical Review Letters}\ }\textbf {\bibinfo {volume} {110}},\ \bibinfo
  {pages} {263201} (\bibinfo {year} {2013})}\BibitemShut {NoStop}%
\bibitem [{Note3()}]{Note3}%
  \BibitemOpen
  \bibinfo {note} {This blockade effect is due to the fact that the vdW
  eigenstates cannot be written as a product state in the uncoupled basis
  [Eq.~\protect \textup {\hbox {\mathsurround \z@ \protect \normalfont
  (\ignorespaces \ref {eq:l1}\unskip \@@italiccorr )}}]}\BibitemShut {NoStop}%
\bibitem [{Note4()}]{Note4}%
  \BibitemOpen
  \bibinfo {note} {For example the symmetric expression of the super pair state
  \protect \textup {\hbox {\mathsurround \z@ \protect \normalfont
  (\ignorespaces \ref {eq:psip}\unskip \@@italiccorr )}} assumes that there is
  no boundary effect.}\BibitemShut {Stop}%
\bibitem [{\citenamefont {Schau\ss}\ \emph {et~al.}(2012)\citenamefont
  {Schau\ss}, \citenamefont {Cheneau}, \citenamefont {Endres}, \citenamefont
  {Fukuhara}, \citenamefont {Hild}, \citenamefont {Omran}, \citenamefont
  {Pohl}, \citenamefont {Gross}, \citenamefont {Kuhr},\ and\ \citenamefont
  {Bloch}}]{Schauss2012}%
  \BibitemOpen
  \bibfield  {author} {\bibinfo {author} {\bibfnamefont {P.}~\bibnamefont
  {Schau\ss}}, \bibinfo {author} {\bibfnamefont {M.}~\bibnamefont {Cheneau}},
  \bibinfo {author} {\bibfnamefont {M.}~\bibnamefont {Endres}}, \bibinfo
  {author} {\bibfnamefont {T.}~\bibnamefont {Fukuhara}}, \bibinfo {author}
  {\bibfnamefont {S.}~\bibnamefont {Hild}}, \bibinfo {author} {\bibfnamefont
  {A.}~\bibnamefont {Omran}}, \bibinfo {author} {\bibfnamefont
  {T.}~\bibnamefont {Pohl}}, \bibinfo {author} {\bibfnamefont {C.}~\bibnamefont
  {Gross}}, \bibinfo {author} {\bibfnamefont {S.}~\bibnamefont {Kuhr}}, \ and\
  \bibinfo {author} {\bibfnamefont {I.}~\bibnamefont {Bloch}},\ }\href
  {\doibase 10.1038/nature11596} {\bibfield  {journal} {\bibinfo  {journal}
  {Nature}\ }\textbf {\bibinfo {volume} {491}},\ \bibinfo {pages} {87}
  (\bibinfo {year} {2012})}\BibitemShut {NoStop}%
\bibitem [{\citenamefont {Malossi}\ \emph {et~al.}(2014)\citenamefont
  {Malossi}, \citenamefont {Valado}, \citenamefont {Scotto}, \citenamefont
  {Huillery}, \citenamefont {Pillet}, \citenamefont {Ciampini}, \citenamefont
  {Arimondo},\ and\ \citenamefont {Morsch}}]{Malossi2014}%
  \BibitemOpen
  \bibfield  {author} {\bibinfo {author} {\bibfnamefont {N.}~\bibnamefont
  {Malossi}}, \bibinfo {author} {\bibfnamefont {M.}~\bibnamefont {Valado}},
  \bibinfo {author} {\bibfnamefont {S.}~\bibnamefont {Scotto}}, \bibinfo
  {author} {\bibfnamefont {P.}~\bibnamefont {Huillery}}, \bibinfo {author}
  {\bibfnamefont {P.}~\bibnamefont {Pillet}}, \bibinfo {author} {\bibfnamefont
  {D.}~\bibnamefont {Ciampini}}, \bibinfo {author} {\bibfnamefont
  {E.}~\bibnamefont {Arimondo}}, \ and\ \bibinfo {author} {\bibfnamefont
  {O.}~\bibnamefont {Morsch}},\ }\href {\doibase
  10.1103/PhysRevLett.113.023006} {\bibfield  {journal} {\bibinfo  {journal}
  {Physical Review Letters}\ }\textbf {\bibinfo {volume} {113}},\ \bibinfo
  {pages} {023006} (\bibinfo {year} {2014})}\BibitemShut {NoStop}%
\bibitem [{\citenamefont {Schempp}\ \emph {et~al.}(2014)\citenamefont
  {Schempp}, \citenamefont {G\"{u}nter}, \citenamefont {Robert-de
  Saint-Vincent}, \citenamefont {Hofmann}, \citenamefont {Breyel},
  \citenamefont {Komnik}, \citenamefont {Sch\"{o}nleber}, \citenamefont
  {G\"{a}rttner}, \citenamefont {Evers}, \citenamefont {Whitlock},\ and\
  \citenamefont {Weidem\"{u}ller}}]{Schempp2014}%
  \BibitemOpen
  \bibfield  {author} {\bibinfo {author} {\bibfnamefont {H.}~\bibnamefont
  {Schempp}}, \bibinfo {author} {\bibfnamefont {G.}~\bibnamefont {G\"{u}nter}},
  \bibinfo {author} {\bibfnamefont {M.}~\bibnamefont {Robert-de
  Saint-Vincent}}, \bibinfo {author} {\bibfnamefont {C.~S.}\ \bibnamefont
  {Hofmann}}, \bibinfo {author} {\bibfnamefont {D.}~\bibnamefont {Breyel}},
  \bibinfo {author} {\bibfnamefont {a.}~\bibnamefont {Komnik}}, \bibinfo
  {author} {\bibfnamefont {D.~W.}\ \bibnamefont {Sch\"{o}nleber}}, \bibinfo
  {author} {\bibfnamefont {M.}~\bibnamefont {G\"{a}rttner}}, \bibinfo {author}
  {\bibfnamefont {J.}~\bibnamefont {Evers}}, \bibinfo {author} {\bibfnamefont
  {S.}~\bibnamefont {Whitlock}}, \ and\ \bibinfo {author} {\bibfnamefont
  {M.}~\bibnamefont {Weidem\"{u}ller}},\ }\href {\doibase
  10.1103/PhysRevLett.112.013002} {\bibfield  {journal} {\bibinfo  {journal}
  {Physical Review Letters}\ }\textbf {\bibinfo {volume} {112}},\ \bibinfo
  {pages} {013002} (\bibinfo {year} {2014})}\BibitemShut {NoStop}%
\bibitem [{\citenamefont {Singer}\ \emph {et~al.}(2005)\citenamefont {Singer},
  \citenamefont {Stanojevic}, \citenamefont {Weidem\"{u}ller},\ and\
  \citenamefont {C\^{o}t\'{e}}}]{Singer2005}%
  \BibitemOpen
  \bibfield  {author} {\bibinfo {author} {\bibfnamefont {K.}~\bibnamefont
  {Singer}}, \bibinfo {author} {\bibfnamefont {J.}~\bibnamefont {Stanojevic}},
  \bibinfo {author} {\bibfnamefont {M.}~\bibnamefont {Weidem\"{u}ller}}, \ and\
  \bibinfo {author} {\bibfnamefont {R.}~\bibnamefont {C\^{o}t\'{e}}},\ }\href
  {\doibase 10.1088/0953-4075/38/2/021} {\bibfield  {journal} {\bibinfo
  {journal} {Journal of Physics B: Atomic, Molecular and Optical Physics}\
  }\textbf {\bibinfo {volume} {38}},\ \bibinfo {pages} {S295} (\bibinfo {year}
  {2005})}\BibitemShut {NoStop}%
\bibitem [{\citenamefont {Marinescu}\ \emph {et~al.}(1994)\citenamefont
  {Marinescu}, \citenamefont {Sadeghpour},\ and\ \citenamefont
  {Dalgarno}}]{Marinescu1994}%
  \BibitemOpen
  \bibfield  {author} {\bibinfo {author} {\bibfnamefont {M.}~\bibnamefont
  {Marinescu}}, \bibinfo {author} {\bibfnamefont {H.}~\bibnamefont
  {Sadeghpour}}, \ and\ \bibinfo {author} {\bibfnamefont {A.}~\bibnamefont
  {Dalgarno}},\ }\href {\doibase 10.1103/PhysRevA.49.982} {\bibfield  {journal}
  {\bibinfo  {journal} {Physical Review A}\ }\textbf {\bibinfo {volume} {49}},\
  \bibinfo {pages} {982} (\bibinfo {year} {1994})}\BibitemShut {NoStop}%
\bibitem [{\citenamefont {Li}\ \emph {et~al.}(2003)\citenamefont {Li},
  \citenamefont {Mourachko}, \citenamefont {Noel},\ and\ \citenamefont
  {Gallagher}}]{Li2003}%
  \BibitemOpen
  \bibfield  {author} {\bibinfo {author} {\bibfnamefont {W.}~\bibnamefont
  {Li}}, \bibinfo {author} {\bibfnamefont {I.}~\bibnamefont {Mourachko}},
  \bibinfo {author} {\bibfnamefont {M.}~\bibnamefont {Noel}}, \ and\ \bibinfo
  {author} {\bibfnamefont {T.}~\bibnamefont {Gallagher}},\ }\href {\doibase
  10.1103/PhysRevA.67.052502} {\bibfield  {journal} {\bibinfo  {journal}
  {Physical Review A}\ }\textbf {\bibinfo {volume} {67}},\ \bibinfo {pages}
  {052502} (\bibinfo {year} {2003})}\BibitemShut {NoStop}%
\bibitem [{\citenamefont {Han}\ \emph {et~al.}(2006)\citenamefont {Han},
  \citenamefont {Jamil}, \citenamefont {Norum}, \citenamefont {Tanner},\ and\
  \citenamefont {Gallagher}}]{Han2006}%
  \BibitemOpen
  \bibfield  {author} {\bibinfo {author} {\bibfnamefont {J.}~\bibnamefont
  {Han}}, \bibinfo {author} {\bibfnamefont {Y.}~\bibnamefont {Jamil}}, \bibinfo
  {author} {\bibfnamefont {D.}~\bibnamefont {Norum}}, \bibinfo {author}
  {\bibfnamefont {P.}~\bibnamefont {Tanner}}, \ and\ \bibinfo {author}
  {\bibfnamefont {T.}~\bibnamefont {Gallagher}},\ }\href {\doibase
  10.1103/PhysRevA.74.054502} {\bibfield  {journal} {\bibinfo  {journal}
  {Physical Review A}\ }\textbf {\bibinfo {volume} {74}},\ \bibinfo {pages}
  {054502} (\bibinfo {year} {2006})}\BibitemShut {NoStop}%
\bibitem [{\citenamefont {Walker}\ and\ \citenamefont
  {Saffman}(2005)}]{Walker2005}%
  \BibitemOpen
  \bibfield  {author} {\bibinfo {author} {\bibfnamefont {T.~G.}\ \bibnamefont
  {Walker}}\ and\ \bibinfo {author} {\bibfnamefont {M.}~\bibnamefont
  {Saffman}},\ }\href {\doibase 10.1088/0953-4075/38/2/022} {\bibfield
  {journal} {\bibinfo  {journal} {Journal of Physics B: Atomic, Molecular and
  Optical Physics}\ }\textbf {\bibinfo {volume} {38}},\ \bibinfo {pages} {S309}
  (\bibinfo {year} {2005})}\BibitemShut {NoStop}%
\bibitem [{\citenamefont {Vaillant}\ \emph {et~al.}(2012)\citenamefont
  {Vaillant}, \citenamefont {Jones},\ and\ \citenamefont
  {Potvliege}}]{Vaillant2012}%
  \BibitemOpen
  \bibfield  {author} {\bibinfo {author} {\bibfnamefont {C.~L.}\ \bibnamefont
  {Vaillant}}, \bibinfo {author} {\bibfnamefont {M.~P.~a.}\ \bibnamefont
  {Jones}}, \ and\ \bibinfo {author} {\bibfnamefont {R.~M.}\ \bibnamefont
  {Potvliege}},\ }\href {\doibase 10.1088/0953-4075/45/13/135004} {\bibfield
  {journal} {\bibinfo  {journal} {Journal of Physics B: Atomic, Molecular and
  Optical Physics}\ }\textbf {\bibinfo {volume} {45}},\ \bibinfo {pages}
  {135004} (\bibinfo {year} {2012})}\BibitemShut {NoStop}%
\end{thebibliography}%

\end{document}